\documentclass[manuscript]{aastex631}

\usepackage{natbib}
\usepackage{multirow}
\usepackage[disable,colorinlistoftodos]{todonotes}
\graphicspath{{./}{figures/}}

\begin{document}

\title{\textit{Insight}-HXMT dedicated 33-day observation of SGR J1935+2154
\uppercase\expandafter{\romannumeral1}. Burst Catalog}

\shorttitle{HXMT SGR J1935+2154 bursts}
\shortauthors{Ce Cai et. al}

\author{Ce Cai\textsuperscript}
\affiliation{Key Laboratory of Particle Astrophysics, Institute of High Energy Physics, Chinese Academy of Sciences, 19B Yuquan Road, Beijing 100049, China}
\affil{University of Chinese Academy of Sciences, Chinese Academy of Sciences, Beijing 100049, China}

\author{Wang-Chen Xue\textsuperscript}
\affiliation{Key Laboratory of Particle Astrophysics, Institute of High Energy Physics, Chinese Academy of Sciences, 19B Yuquan Road, Beijing 100049, China}
\affil{University of Chinese Academy of Sciences, Chinese Academy of Sciences, Beijing 100049, China}

\author{Cheng-Kui Li\textsuperscript{*}}
\email{lick@ihep.ac.cn}
\affiliation{Key Laboratory of Particle Astrophysics, Institute of High Energy Physics, Chinese Academy of Sciences, 19B Yuquan Road, Beijing 100049, China}

\author{Shao-Lin Xiong\textsuperscript{*}}
\email{xiongsl@ihep.ac.cn}
\affiliation{Key Laboratory of Particle Astrophysics, Institute of High Energy Physics, Chinese Academy of Sciences, 19B Yuquan Road, Beijing 100049, China}

\author{Shuang-Nan Zhang\textsuperscript{*}}
\email{zhangsn@ihep.ac.cn}
\affiliation{Key Laboratory of Particle Astrophysics, Institute of High Energy Physics, Chinese Academy of Sciences, 19B Yuquan Road, Beijing 100049, China}

\author{Lin Lin}
\affiliation{Department of Astronomy, Beijing Normal University, Beijing 100088, People’s Republic of China}

\author{Xiao-Bo Li}
\affiliation{Key Laboratory of Particle Astrophysics, Institute of High Energy Physics, Chinese Academy of Sciences, 19B Yuquan Road, Beijing 100049, China}

\author{Ming-Yu Ge}
\affiliation{Key Laboratory of Particle Astrophysics, Institute of High Energy Physics, Chinese Academy of Sciences, 19B Yuquan Road, Beijing 100049, China}

\author{Hai-Sheng Zhao}
\affiliation{Key Laboratory of Particle Astrophysics, Institute of High Energy Physics, Chinese Academy of Sciences, 19B Yuquan Road, Beijing 100049, China}

\author{Li-Ming Song}
\affiliation{Key Laboratory of Particle Astrophysics, Institute of High Energy Physics, Chinese Academy of Sciences, 19B Yuquan Road, Beijing 100049, China}
\affiliation{University of Chinese Academy of Sciences, Chinese Academy of Sciences, Beijing 100049, China}

\author{Fang-Jun Lu}
\affiliation{Key Laboratory of Particle Astrophysics, Institute of High Energy Physics, Chinese Academy of Sciences, 19B Yuquan Road, Beijing 100049, China}
\affiliation{University of Chinese Academy of Sciences, Chinese Academy of Sciences, Beijing 100049, China}

\author{Shu Zhang}
\affiliation{Key Laboratory of Particle Astrophysics, Institute of High Energy Physics, Chinese Academy of Sciences, 19B Yuquan Road, Beijing 100049, China}

\author{Yan-Qiu Zhang}
\affiliation{Key Laboratory of Particle Astrophysics, Institute of High Energy Physics, Chinese Academy of Sciences, 19B Yuquan Road, Beijing 100049, China}
\affiliation{University of Chinese Academy of Sciences, Chinese Academy of Sciences, Beijing 100049, China}

\author{Shuo Xiao}
\affiliation{Key Laboratory of Particle Astrophysics, Institute of High Energy Physics, Chinese Academy of Sciences, 19B Yuquan Road, Beijing 100049, China}
\affiliation{University of Chinese Academy of Sciences, Chinese Academy of Sciences, Beijing 100049, China}

\author{You-Li Tuo}
\affiliation{Key Laboratory of Particle Astrophysics, Institute of High Energy Physics, Chinese Academy of Sciences, 19B Yuquan Road, Beijing 100049, China}

\author{Qi-Bin Yi}
\affiliation{Key Laboratory of Particle Astrophysics, Institute of High Energy Physics, Chinese Academy of Sciences, 19B Yuquan Road, Beijing 100049, China}
\affiliation{School of Physics and Optoelectronics, Xiangtan University, Xiangtan 411105, Hunan, China}

\author{Zhi Wei Guo}
\affiliation{Key Laboratory of Particle Astrophysics, Institute of High Energy Physics, Chinese Academy of Sciences, 19B Yuquan Road, Beijing 100049, China}
\affiliation{College of physics Sciences Technology, Hebei University, No. 180 Wusi Dong Road, Lian Chi District, Baoding City, Hebei Province 071002, China}

\author{Sheng Lun Xie}
\affiliation{Key Laboratory of Particle Astrophysics, Institute of High Energy Physics, Chinese Academy of Sciences, 19B Yuquan Road, Beijing 100049, China}
\affiliation{School of Physical Science and Technology, Central China Normal University, Wuhan 430097, China}

\author{Yi Zhao}
\affiliation{Key Laboratory of Particle Astrophysics, Institute of High Energy Physics, Chinese Academy of Sciences, 19B Yuquan Road, Beijing 100049, China}
\affiliation{Department of Astronomy, Beijing Normal University, Beijing 100088, People’s Republic of China}

\author{Zhen Zhang}
\affiliation{Key Laboratory of Particle Astrophysics, Institute of High Energy Physics, Chinese Academy of Sciences, 19B Yuquan Road, Beijing 100049, China}

\author{Qing-Xin Li}
\affiliation{Department of Astronomy, Beijing Normal University, Beijing 100088, People’s Republic of China}

\author{Jia-Cong Liu}
\affiliation{Key Laboratory of Particle Astrophysics, Institute of High Energy Physics, Chinese Academy of Sciences, 19B Yuquan Road, Beijing 100049, China}
\affiliation{University of Chinese Academy of Sciences, Chinese Academy of Sciences, Beijing 100049, China}

\author{Chao Zheng}
\affiliation{Key Laboratory of Particle Astrophysics, Institute of High Energy Physics, Chinese Academy of Sciences, 19B Yuquan Road, Beijing 100049, China}
\affiliation{University of Chinese Academy of Sciences, Chinese Academy of Sciences, Beijing 100049, China}

\author{Ping Wang}
\affiliation{Key Laboratory of Particle Astrophysics, Institute of High Energy Physics, Chinese Academy of Sciences, 19B Yuquan Road, Beijing 100049, China}

\begin{abstract}
Magnetars are neutron stars with extreme magnetic field and sometimes manifest as soft gamma-ray repeaters (SGRs). SGR J1935+2154 is one of the most prolific bursters and the first confirmed source of fast radio burst (i.e. FRB 200428).
Encouraged by the discovery of the first X-ray counterpart of FRB, \textit{Insight}-Hard X-ray Modulation Telescope (\textit{Insight}-HXMT) implemented a dedicated 33-day long ToO observation of SGR J1935+2154 since April 28, 2020. With the HE, ME and LE telescopes, \textit{Insight}-HXMT provides a thorough monitoring of burst activity evolution of SGR J1935+2154, in a very broad energy range (1-250 keV) with high temporal resolution and high sensitivity, resulting in a unique valuable data set for detailed studies of SGR J1935+2154.
In this work, we conduct a comprehensive analysis of this observation including detailed burst search, identification and temporal analyses.
After carefully removing false triggers, we find a total of 75 bursts from SGR J1935+2154, out of which 70 are single-pulsed. The maximum burst rate is about 56 bursts/day. Both the burst duration and the waiting time between two successive bursts follow log-normal distributions, consistent with previous studies.
We also find that bursts with longer duration (some are multi-pulsed) tend to occur during the period with relatively high burst rate.
There is no correlation between the waiting time and the fluence or duration of either the former or latter burst. It also seems that there is no correlation between burst duration and hardness ratio, in contrast to some previous reports.
In addition, we do not find any X-ray burst associated with any reported radio bursts except for FRB 200428.
\end{abstract}


\section{Introduction}
Soft gamma-ray repeaters (SGRs), manifestation of celestial objects unpredictably emitting bursts in hard X-rays and soft gamma-rays, are deemed to be rotating neutron stars with ultra-strong surface magnetic fields up to 10$^{14}$ $-$ 10$^{15}$ G, known as magnetars \citep{1995MNRAS.275..255T,1998Natur.393..235K}. These magnetar bursts are most likely powered by the release of the magnetic field energy triggered by the starquakes of neutron stars \citep{1995MNRAS.275..255T} or the reconnection of the magnetic field lines \citep{2003MNRAS.346..540L}.

Magnetars are also plausible sources of fast radio bursts (FRBs), which are another kind of mysterious cosmological flashes in the radio band. This connection was first built on the sufficient energy preserved by the strong magnetic field of magnetars and the compatible rate of magnetar bursts in X-rays with that of FRBs \citep{2010vaoa.conf..129P}. Recently it has been proved by observations of a bursting magnetar SGR J1935+2154.

SGR J1935+2154 is one of the most prolific magnetars in our galaxy, which first triggered the Burst Alert Telescope (BAT) aboard the Neil Gehrels Swift Observatory (hereafter \textit{Swift}) through a short hard X-ray burst on July 5th, 2014 \citep{2014GCN.16520....1S}. Since then, SGR J1935+2154 has been sporadically active, with several outburst episodes happened in the years of 2014, 2015, 2016, 2019 and 2020 \citep{2017ApJ...847...85Y,Lin_2020,Lin_2020b}. On April 27th, 2020, SGR J1935+2154 entered an active phase with the emission of hundreds of bursts.
Later, a very bright fast radio burst (FRB~200428) was detected by CHIME and STARE2 from the direction of the bursting magnetar \citep{2020Natur.587...54C,bochenek2020fast}.


This FRB is found to be associated with an unusual non-thermal X-ray burst accurately localized to SGR J1935+2154 by \textit{Insight}-HXMT \citep{2021NatAs...5..378L} and INTEGRAL \citep{Mereghetti_2020} and jointly detected by other gamma-ray detectors including AGILE \citep{tavani2020xray} and Konus-Wind \citep{2021NatAs...5..372R}, providing an unambiguous proof that SGR J1935+2154 is the first confirmed source that can radiate the FRB.

\textit{Insight}-HXMT is China’s first X-ray astronomy satellite \citep{Zhang_2020,Shu_2018}, launched on June 15th, 2017. It is composed of three telescopes with different energy bands: the High Energy X-ray telescope (HE, 20$–$250 keV) \citep{2019arXiv191004955L}, the Medium Energy X-ray telescope (ME, 5$–$30 keV) \citep{2019arXiv191004451C} and the Low Energy X-ray telescope (LE, 1$–$15 keV) \citep{2019arXiv191008319C}. With wide energy band and high time resolution, \textit{Insight}-HXMT plays an important role in searching for the X-ray and gamma-ray transient, with duration from milliseconds to seconds, as gamma-ray bursts (GRBs) \citep{10.1093/mnras/stab2760}, high-energy emission of FRBs \citep{Guidorzi_2020,2020Guidorzi} and electromagnetic counterpart of gravitational waves \citep{Li:2017iup, Zhang_2020, 10.1093/mnras/stab2760}.

\textit{Insight}-HXMT discovered an X-ray burst associated with FRB 200428 (called FRB 200428-Associated Burst hereafter) from the Galactic magnetar SGR J1935+2154 \citep{2021NatAs...5..378L}. We first suggested that there are two narrow peaks in this X-ray burst that are the high energy counterparts of the two radio peaks in FRB 200428 \citep{2020Natur.587...54C,bochenek2020fast,2021NatAs...5..378L}. This breakthrough discovery encouraged us to implement a dedicated month long Target of Opportunity (ToO) observations to continuously monitor SGR J1935+2154, to give a substantially deep depiction of the burst activity evolution of this magnetar. During this long ToO, the first X-ray bursts of SGR J1935+2154 detected by \textit{Insight}-HXMT is about 13 hours after the onset of burst activity, which is marked by the very first burst detected by \textit{Fermi}/GBM and \textit{Swift}/BAT at 2020-04-27T18:26:20 UTC \citep{2020GCN.27665....1S,2020GCN.27659....1F,2020GCN.27660....1P}.

Thanks to the wide energy band and high sensitivity of \textit{Insight}-HXMT, this ToO observation attain a unique sample of bursts in 1 keV to 250 keV with a flux lower limit of about 1 $\times 10^{-9}$ erg cm$^{-2}$ s$^{-1}$ \citep{2020ATel13729....1Z}, which significantly enlarge the burst samples of SGR J1935+2154 detected by other instruments (e.g., \textit{Swift} \citep{2004ApJ...611.1005G}, \textit{NICER} \citep{2016SPIE.9905E..1HG}, \textit{Fermi}/GBM \citep{2009ApJ...702..791M}) during this interesting burst episode with the emission of a FRB.

With this \textit{Insight}-HXMT burst sample, we implement a series of detailed data analyses. As the paper \uppercase\expandafter{\romannumeral1} of this paper series for \textit{Insight}-HXMT observation of SGR J1935+2154, we focus on the burst search, classification and verification of bursts, and temporal analyses. In paper \uppercase\expandafter{\romannumeral2}, we will give the detailed time-integrated spectral analysis of these bursts from SGR J1935+2154.

This paper is organized as follows: In section \ref{sec:Instrumentation_SAMPLES}, we summarize the observations and data reduction. Section \ref{sec:ANALYSIS_METHOD} describes the trigger algorithm, classification and verification of bursts. We present the catalog analysis method and results in section \ref{sec:result}. Finally, we give the discussion and summary in Section \ref{sec:discussion} and Section \ref{sec:summary}.

\begin{table*}[htbp]
\begin{center}
\caption{\textit{Insight}-HXMT observations of SGR J1935+2154 and Blank Sky (for background study).}
\begin{tabular}{ p{3cm}<{\centering}  p{2.5cm}<{\centering} p{3.5cm}<{\centering} p{3.5cm}<{\centering} p{3cm}<{\centering}} 
\hline
\hline
Target Name & obsID & Start Time$^{1}$ (UTC) & End Time$^{1}$ (UTC) & Exposure$^{1}$ (ks) \\
\hline
SGR J1935+2154 $^{2}$
    & P0314003001 &2020-04-28T07:14:51.000  &2020-04-29T12:02:36.000   & 60   \\
SGR J1935+2154 $^{2}$    & P0314003002 &2020-04-30T06:58:23.000  & 2020-05-06T01:20:55.000  & 290   \\
SGR J1935+2154 $^{2}$    & P0314003003 &2020-05-06T01:20:55.000  & 2020-05-08T01:03:20.000  & 100   \\
SGR J1935+2154 $^{2}$    & P0314005$^{4}$ &2020-05-08T01:03:20.000  & 2020-06-01T00:00:01.000  & 1200  \\
\hline      
Blank Sky $^{3}$ & P0101293$^{5}$ &2017-11-02T05:00:57.000  & 2018-03-27T05:24:47.000  & 1796 \\
\hline 
\end{tabular}
\tablecomments{
$^1$ Start time, end time and total exposure time of each observation.\\
$^2$ HE, ME and LE have exposures of 1650 ks, 1479 ks and 1339 ks for SGR J1935+2154 observations respectively. \\
$^3$ HE, ME and LE have exposures of 1796 ks, 1339 ks and 1123 ks for blank sky observations used in this work.\\
$^4$ Includes 8 observations from P0314005001 to P0314005008.\\
$^5$ Includes many observations starting with P0101293001. \\}
\label{tab:table1}
\end{center}
\end{table*}

\section{Observation and Data SAMPLES}
 \label{sec:Instrumentation_SAMPLES}

Since 2020-04-28T07:14:51 UTC, \textit{Insight}-HXMT started the dedicated 33-day (a total span time of 2851.2 ks) ToO observations of SGR J1935+2154, which had emitted hundreds of short X-ray bursts and triggered a series of astronomical satellites starting from 2020 April 27 \citep{2020GCN.27660....1P}. This \textit{Insight}-HXMT dedicated long ToO ended at 2020-06-01T00:00:01 UTC with a total effective exposure of 1650 ks, and covered the time of FRB 200428. The detailed observation time history of \textit{Insight}-HXMT is listed in Table \ref{tab:table1}.

To facilitate the joint observations of SGR J1935+2154 with multi-wavelength telescopes, a preliminary search of X-ray bursts has been done and a preliminary list of X-ray bursts has been released on the \textit{Insight}-HXMT  website\footnote{http://hxmtweb.ihep.ac.cn/bursts/392.jhtml}. In the present work, we refine the burst search for SGR J1935+2154 X-ray bursts and implement a comprehensive analysis with the refined bursts sample, with the data acquired from the telescopes of HE, ME and LE of \textit{Insight}-HXMT. We obtain the screened event files with high time resolution by analysing the 1K data of HE, ME and LE with the up-to-date version (version 2.04) of \textit{Insight}-HXMT Data Analysis Software package (hereafter\textit{HXMTDAS}) \footnote{http://www.hxmt.cn/}.

First, the raw event files are processed to produce calibrated event files \citep{2020JHEAp..27...64L}. We use the commands \textit{hepical}, \textit{mepical} and \textit{lepical} to calibrate the photon events of HE, ME and LE, respectively. The HE command \textit{hepical} is also used to remove the spike events, which are caused by the interactions of high energy cosmic rays with satellite materials and usually occurred simultaneously in a few NaI and CsI detectors \citep{wu2022removal}. The ME command \textit{megrade} is used to calculate the event grade and deadtime of each Field Programmable Gate Array (FPGA). The LE command \textit{lerecon} is used to reconstruct two spilt events and assign event grades.

The commands of \textit{hegtigen}, \textit{megtigen} and \textit{legtigen} are used to select the good time intervals (GTIs). The ELV and SAA flag parameters (“ELV$>$1” and “SAA flag=0”) are set to exclude the time intervals when the satellite flies through SAA or the targeted source is blocked by the Earth.
In addition, the \textit{megticorr} and \textit{legticorr} commands are used to generate new GTI files to further eliminate some bad time intervals, including some events with higher grades, which are not easy to calibrate.

Finally, the commands \textit{hescreen}, \textit{mescreen} and \textit{lescreen} are used to extract the good events using the GTIs above.
The screened event files including these good events are used to search for X-ray bursts down to ms timescales. We use all un-blinded detectors of LE, ME and NaI detectors of HE (CsI detectors of HE is not used since the spectra of SGR bursts are not hard enough to leave signals in CsI crystals). The selected energy bands of HE, ME, LE are 28$-$250 keV, 10$-$30 keV and 1$-$10 keV, respectively.

\begin{table*}
\begin{center}
\caption{Trigger algorithms and criteria used in the burst joint search of SGR J1935+2154 and the test results of the search on the blank sky data.}
\begin{tabular}{p{2.1cm}<{\centering} p{1.7cm}<{\centering} p{1.9cm} p{1cm} p{0.8cm} p{0.8cm} p{0.8cm}  p{1.5cm} p{0.8cm} p{1.2cm} p{0.6cm}} 
\hline
\hline
Timescales (s) & Phases (s) & Telescopes & ${C}_{\rm HE}^{1}$ & ${N}_{\rm HE}^{2}$ & ${C}_{\rm ME}^{3}$ & ${C}_{\rm LE}^{3}$ &  GTIs$^{4}$ (s) & ${N_{\rm FT}^{5}}$ & ${P_{\rm FA}^{6}}$  & Sig$^{7}$\\
\hline
0.005   & 0/0.0025   & HE\&ME  & 1 & 4 & 3 & - &  1.33$\times 10^6$ & 0 & - &  -\\
0.005   & 0/0.0025      & HE\&LE  & 1 & 4 & - & 3 &  1.12$\times 10^6$ & 1 & 4.45e-09 & 5.7$\sigma$ \\
0.005   & 0/0.0025       & ME\&LE  & - & - & 5 & 5 &  1.12$\times 10^6$ & 0 & - & - \\
0.005   & 0/0.0025       & HE\&ME\&LE &1 & 3 & 1 & 1 &  1.12$\times 10^6$ & 1 & 4.45e-09 & 5.7$\sigma$\\

 \hline
0.01  & 0/0.005     & HE\&ME  & 2 & 5 & 2 & - &  1.33$\times 10^6$ & 0 & - &  -  \\
0.01  & 0/0.005       & HE\&LE  & 2 & 5 & - & 2 &  1.12$\times 10^6$ & 0 & - &  - \\
0.01  & 0/0.005     & ME\&LE  & - & - & 5 & 5 &  1.12$\times 10^6$ & 9 & 8.01e-08  & 5.2$\sigma$ \\
0.01  & 0/0.005      & HE\&ME\&LE  & 2 & 1 & 3 & 3 &  1.12$\times 10^6$ & 1 & 8.90e-09  & 5.6$\sigma$ \\
    
 \hline
0.02  & 0/0.01     & HE\&ME  & 3 & 3 & 3 & - &  1.33$\times 10^6$ & 0 & -  & - \\
0.02  & 0/0.01        & HE\&LE  & 3 & 3 & - & 3 &  1.12$\times 10^6$ & 1 & 1.78e-08 &5.51$\sigma$  \\
0.02  & 0/0.01        & HE\&ME\&LE  & 3 & 1 & 3 & 3 &1.12$\times 10^6$&6& 1.06e-07  & 5.18$\sigma$  \\
    
 \hline
0.04  & 0/0.02     & HE\&ME  & 3 & 3 & 8 & -  & 1.33$\times 10^6$ & 0 & -  & - \\
0.04  & 0/0.02       & HE\&LE  & 3 & 3 & - &8 & 1.12$\times 10^6$ & 14 & 4.98e-07 & 4.89$\sigma$  \\
0.04  & 0/0.02   & HE\&ME\&LE  & 3 & 1 & 8 & 8  & 1.12$\times 10^6$ & 0 & -  & - \\
\hline 
\end{tabular}
\tablecomments{$^1$ The minimum net counts of each NaI detector of HE required for a trigger. \\
$^2$ The minimum number of HE detectors required for a trigger. \\
$^3$ The minimum net counts for all detectors of ME or LE required for a trigger. \\
$^4$ Common good time intervals of different telescopes for the blank sky observations.\\
$^5$ Number of false triggers (i.e. FT) found in the searching of the blank sky data.\\
$^6$ False alarm probability derived (formula \ref{FAP}) from the search results of the blank sky data. \\
$^7$ The equivalent Gaussian significance of the false alarm probability.\\}
\label{tab:table2}
\end{center}
\end{table*}

\section{BURST SEARCH and IDENTIFICATION} \label{sec:ANALYSIS_METHOD}

\begin{figure}
\centering
\begin{tabular}{c}
\includegraphics[width=0.50\textwidth]{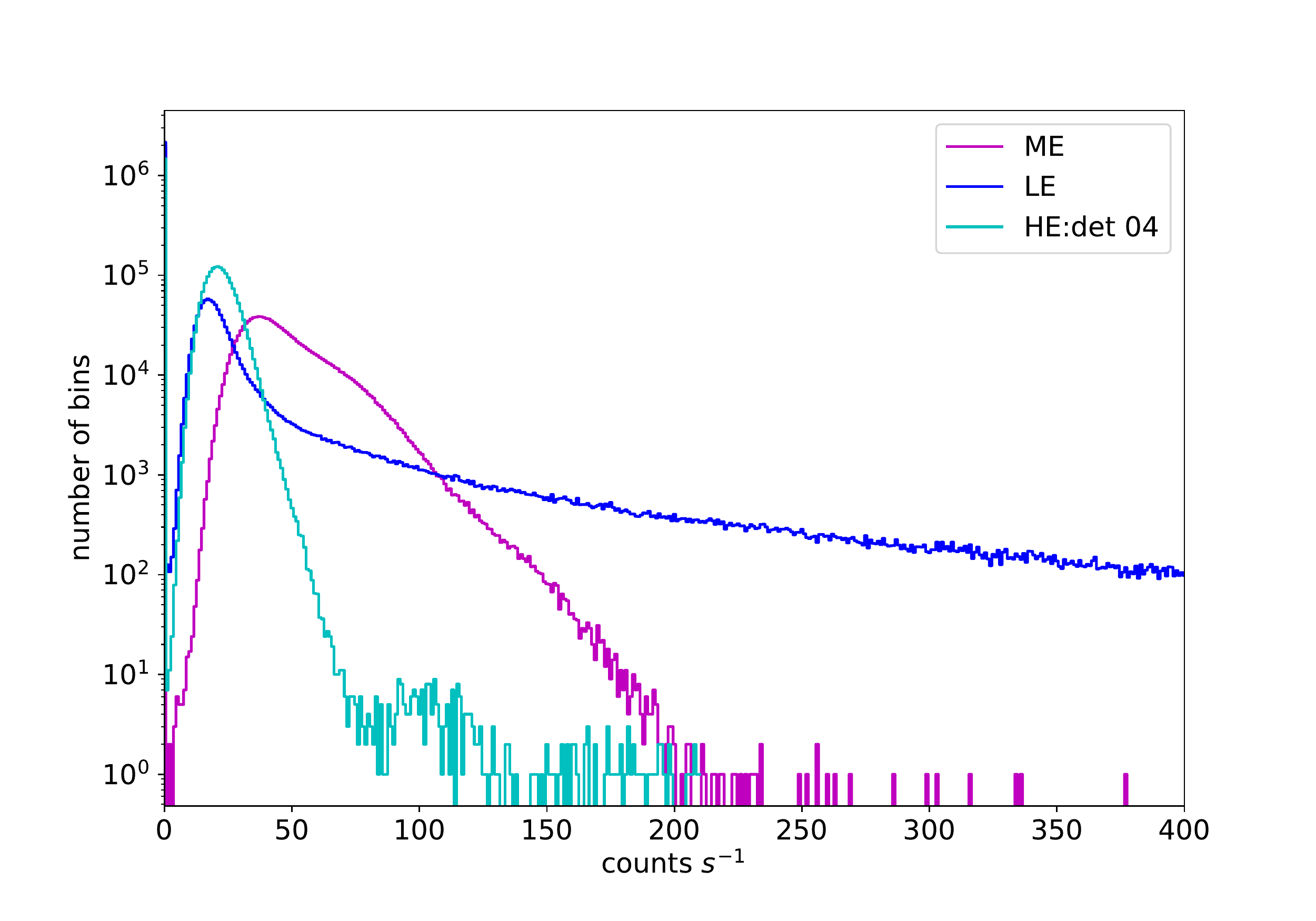}  \\
\end{tabular}
\caption{\label{fig:BackgroundDis} The count rate (counts s$^{-1}$ ) distribution of the background of the three telescopes of \textit{Insigth}-HXMT during blank sky observations. The blue and purple lines represent LE (1$-$10 keV) and ME (10$-$30 keV) respectively. The green line is for NaI detector \#04 of HE (28$-$250 keV), of which the counts of spikes are tried to be removed by \textit{HXMTDAS}.
NaI detector \#04 could represent other NaI detectors of HE. The HE detectors are denoted as 00$-$17 (detector \# 16 is the blinded detector which is not used in this work).}
\end{figure}

\begin{figure}
\centering
\begin{tabular}{c}
\includegraphics[width=0.50\textwidth]{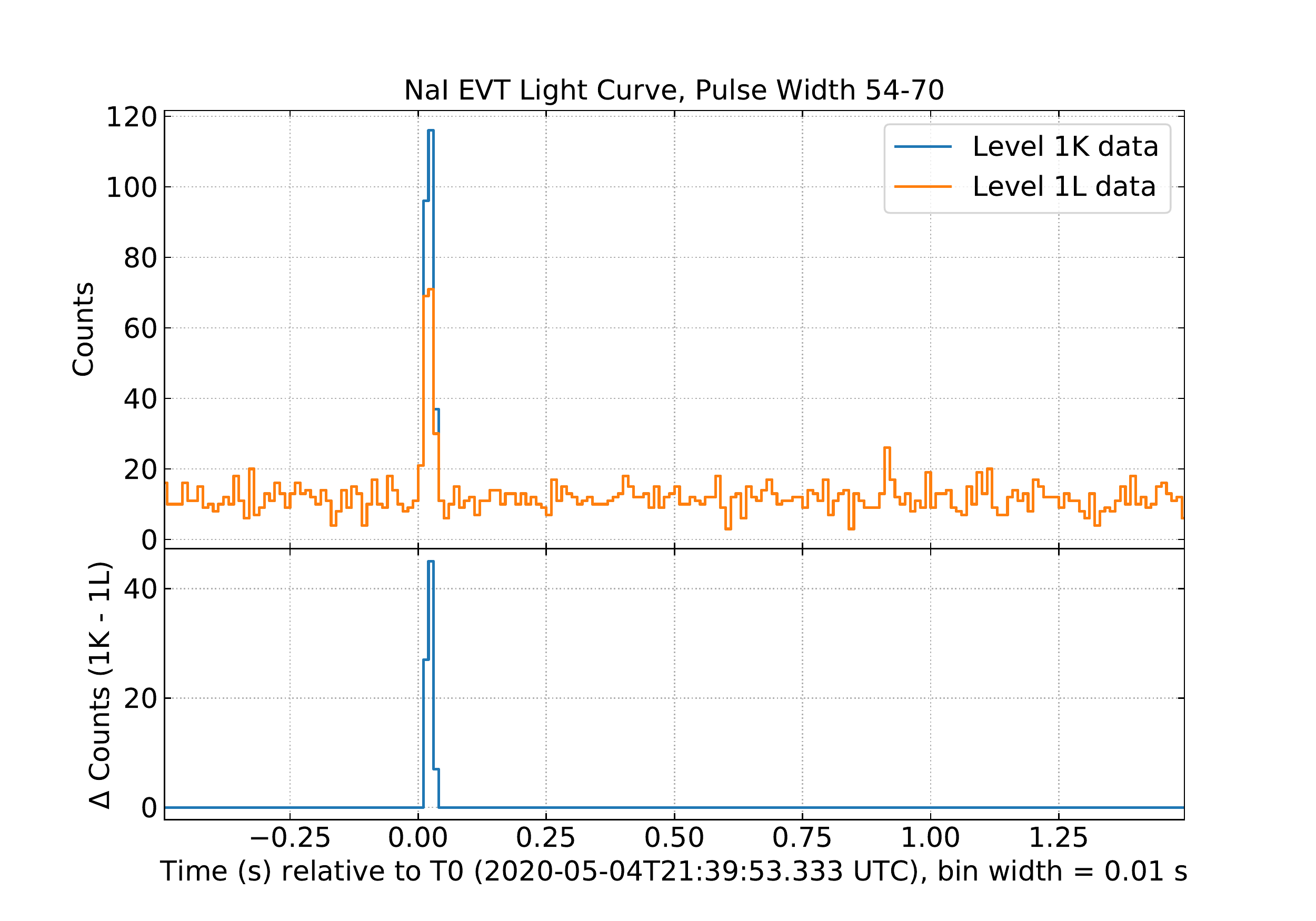}  \\
\end{tabular}
\caption{\label{fig:spike_filter}
\textit{Top}: Summed light curves of all NaI detectors for a burst candidate which was included in the preliminary bursts list but identified as spikes in this work. $T_0$ is the trigger time. The blue line is the light curve made with raw data without removing spikes (i.e. Level 1K data), while the orange line is the light curve made with the Level 1L data for which the counts of spikes were tried to be filtered out by \textit{HXMTDAS}. However, the residual spike counts are still clearly seen. \textit{Bottom}: The difference between the raw light curve and the light curve after removing spikes shown in the top panel.}
\end{figure}

\begin{figure*}
\centering
\begin{tabular}{cc}
\includegraphics[width=0.50\textwidth]{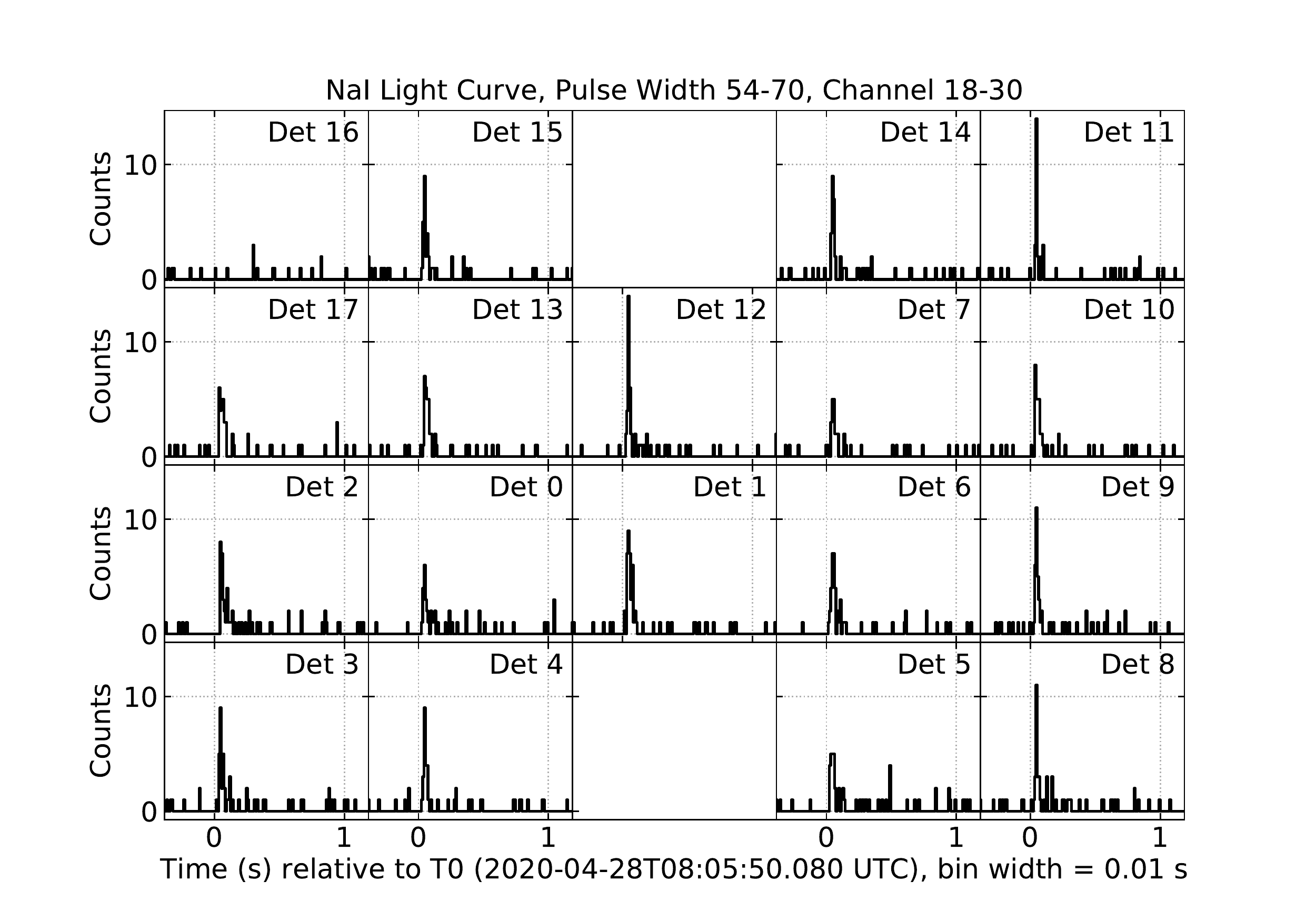} &
\includegraphics[width=0.50\textwidth]{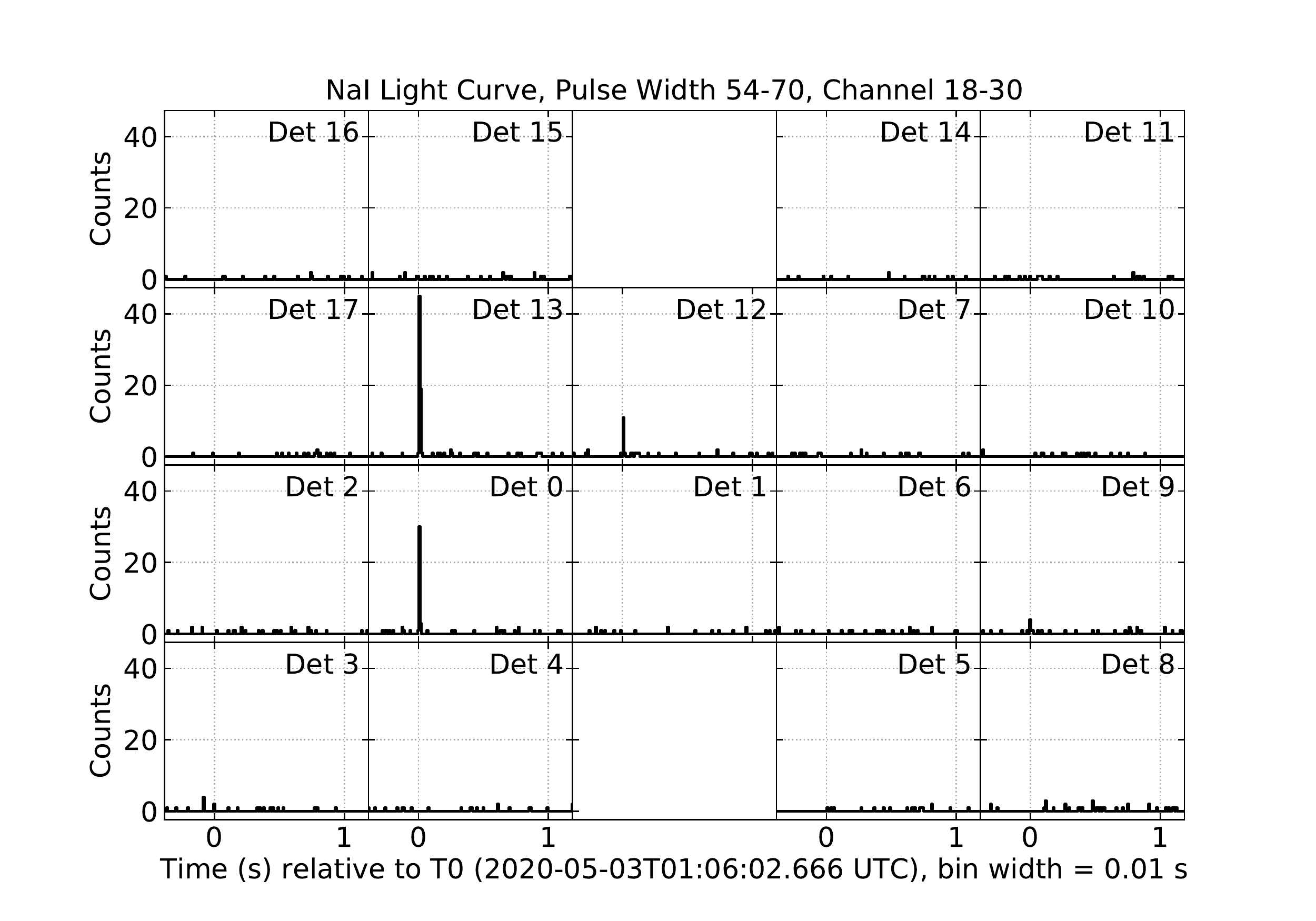}  \\
\end{tabular}
\caption{\label{fig:spike_burst} The light curves of 18 NaI detectors of HE. \textit{Left}:  $T_0$ is the trigger time of a burst candidate which is identified as the SGR J1935+2154 burst. Count increases are seen in all detectors, except for the blinded detector (det \#16). \textit{Right}: Light curves for a burst candidate which is identified as a spike in this work. The excess counts are only seen in a few adjacent detectors which is the characteristics of spikes.}
\end{figure*}

\begin{center}
\begin{figure*}
	\includegraphics[width=60mm]{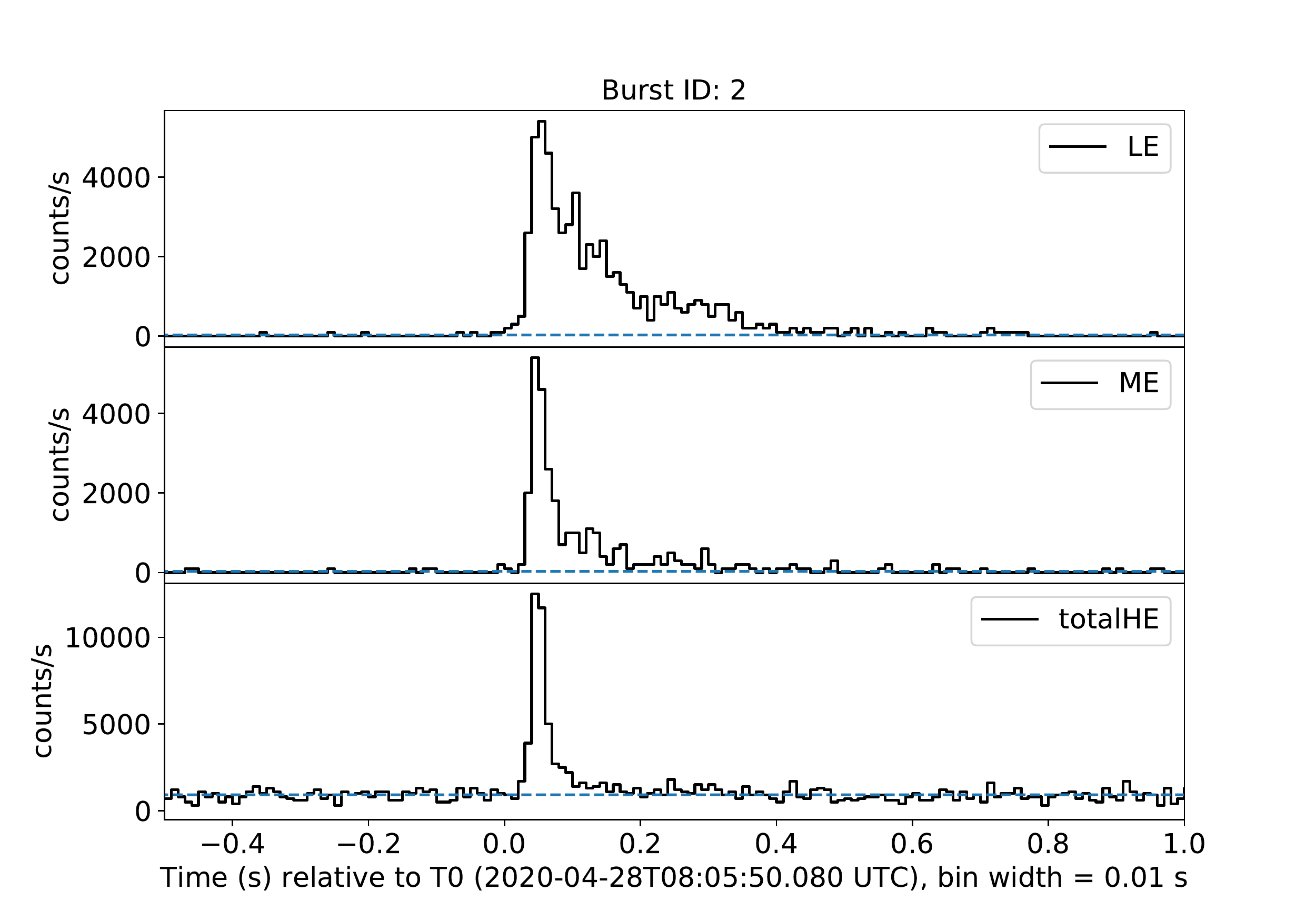}
	\includegraphics[width=60mm]{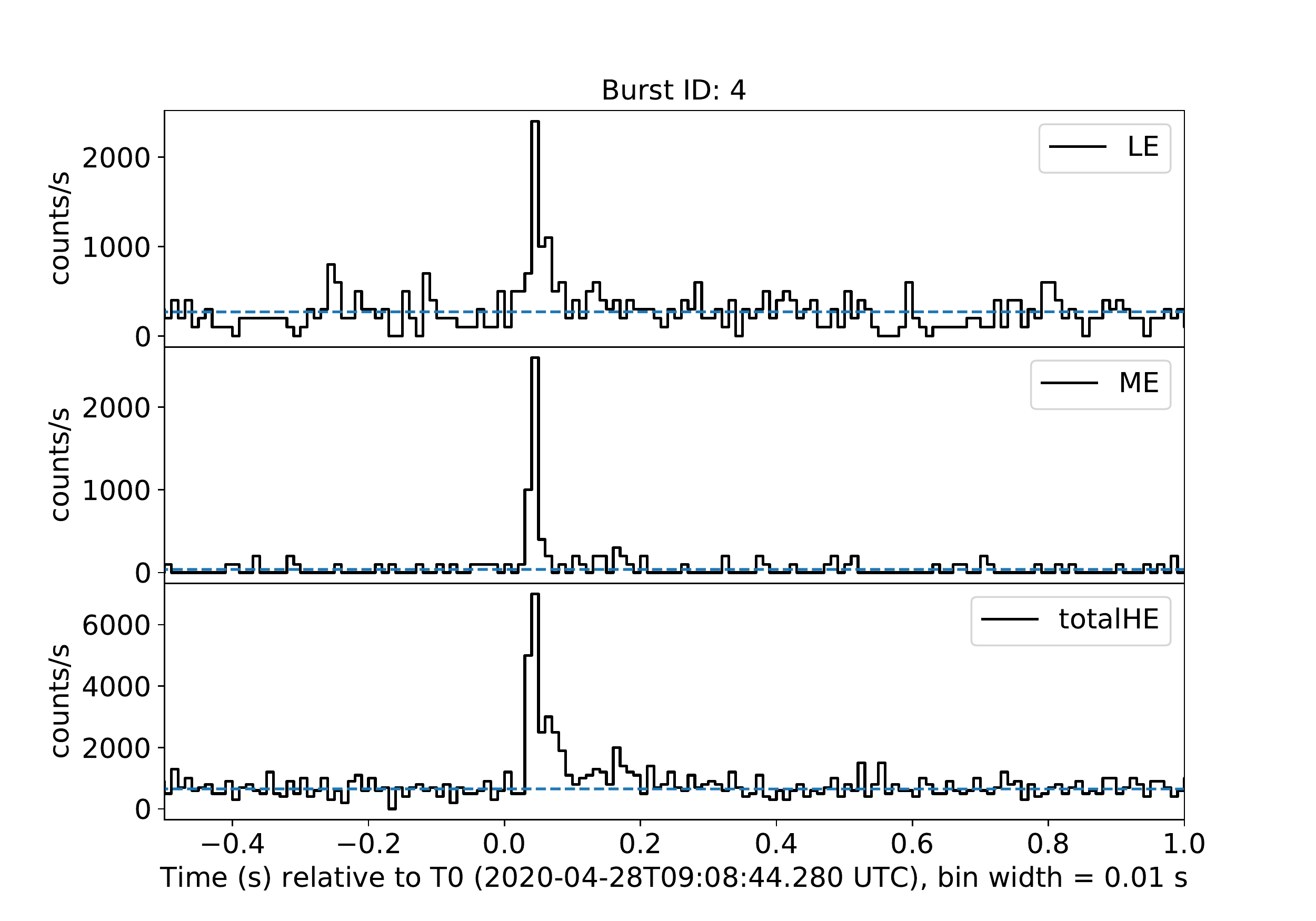}
	\includegraphics[width=60mm]{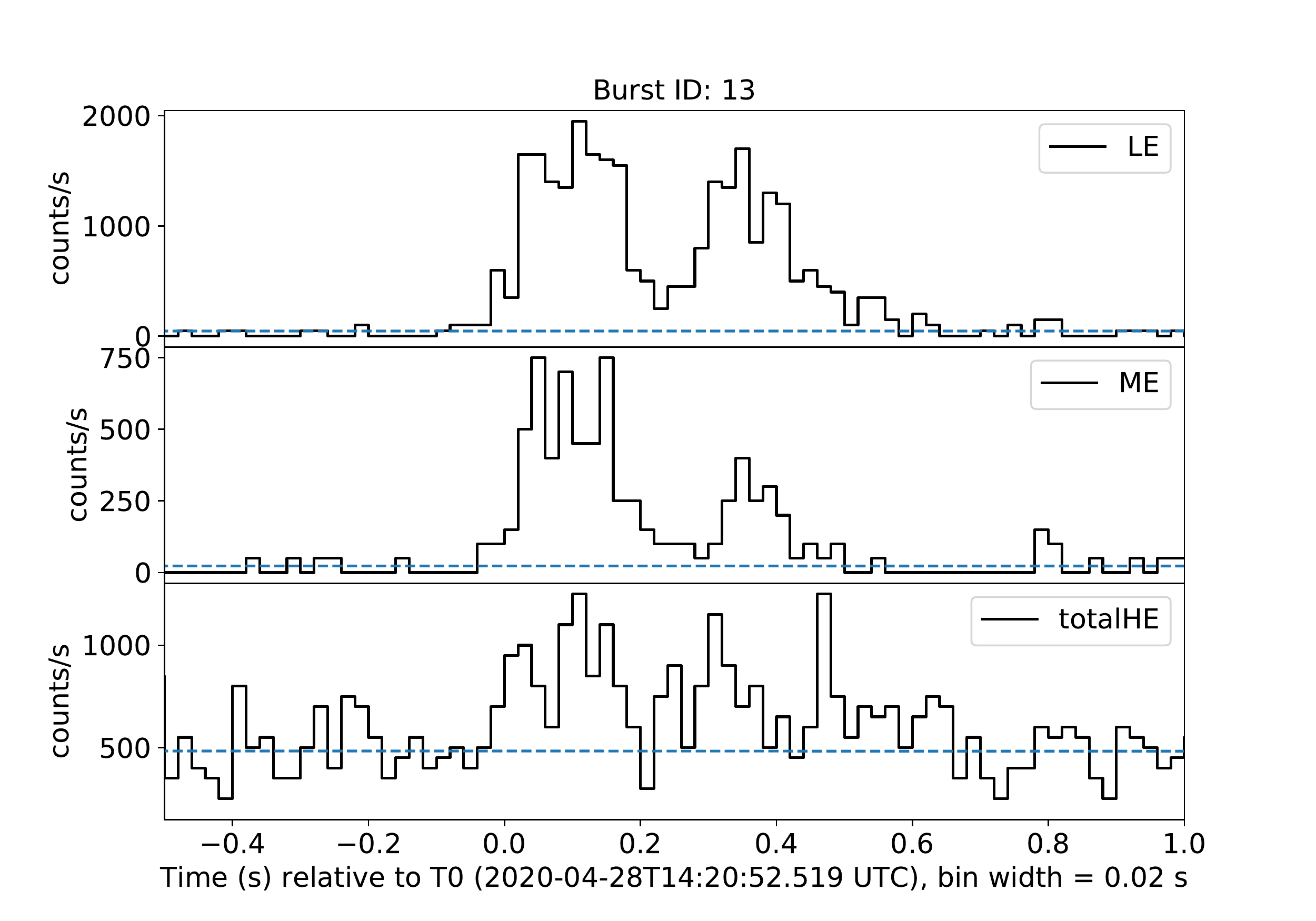}

    \vspace{1mm}

	\includegraphics[width=60mm]{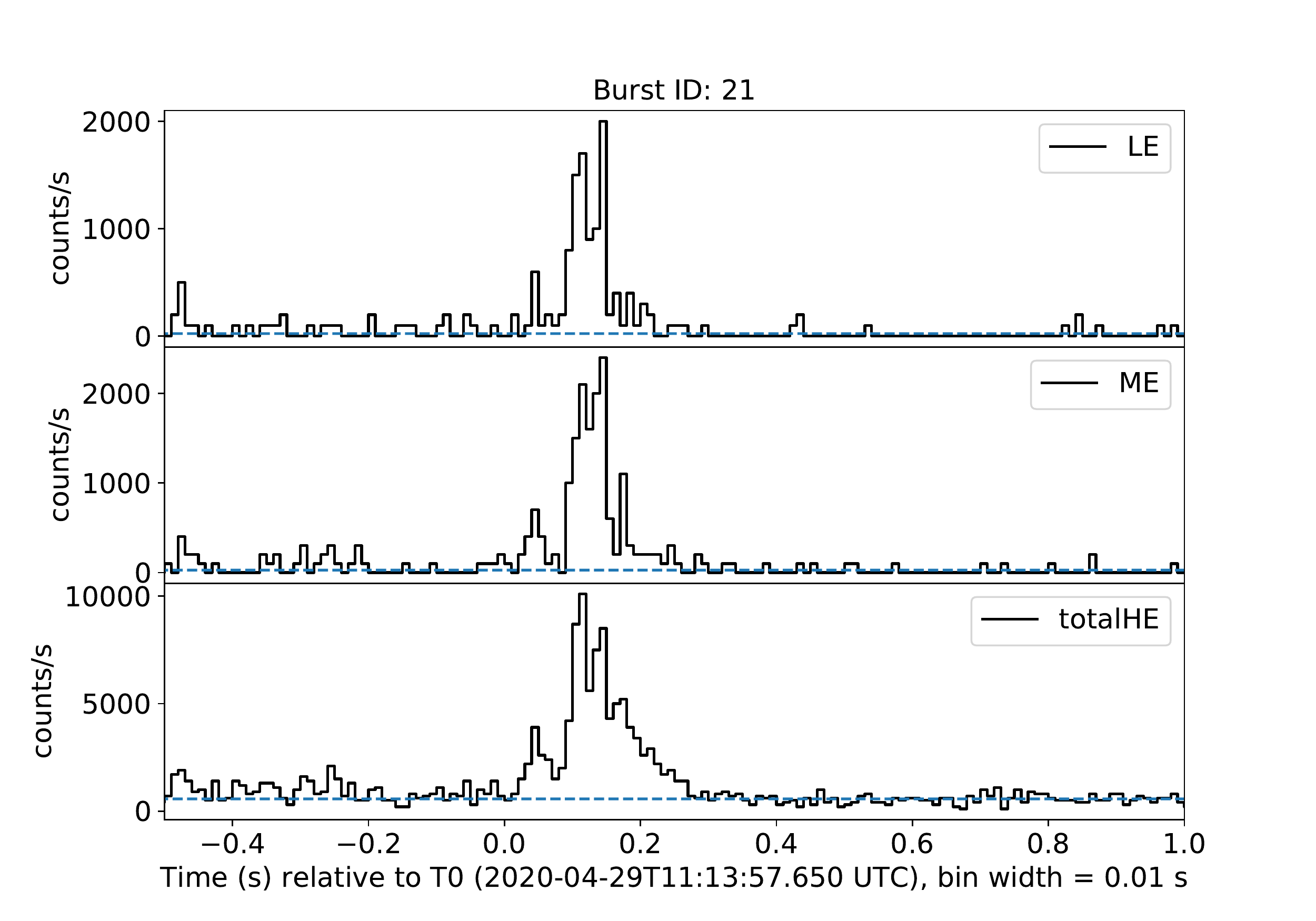}
	\includegraphics[width=60mm]{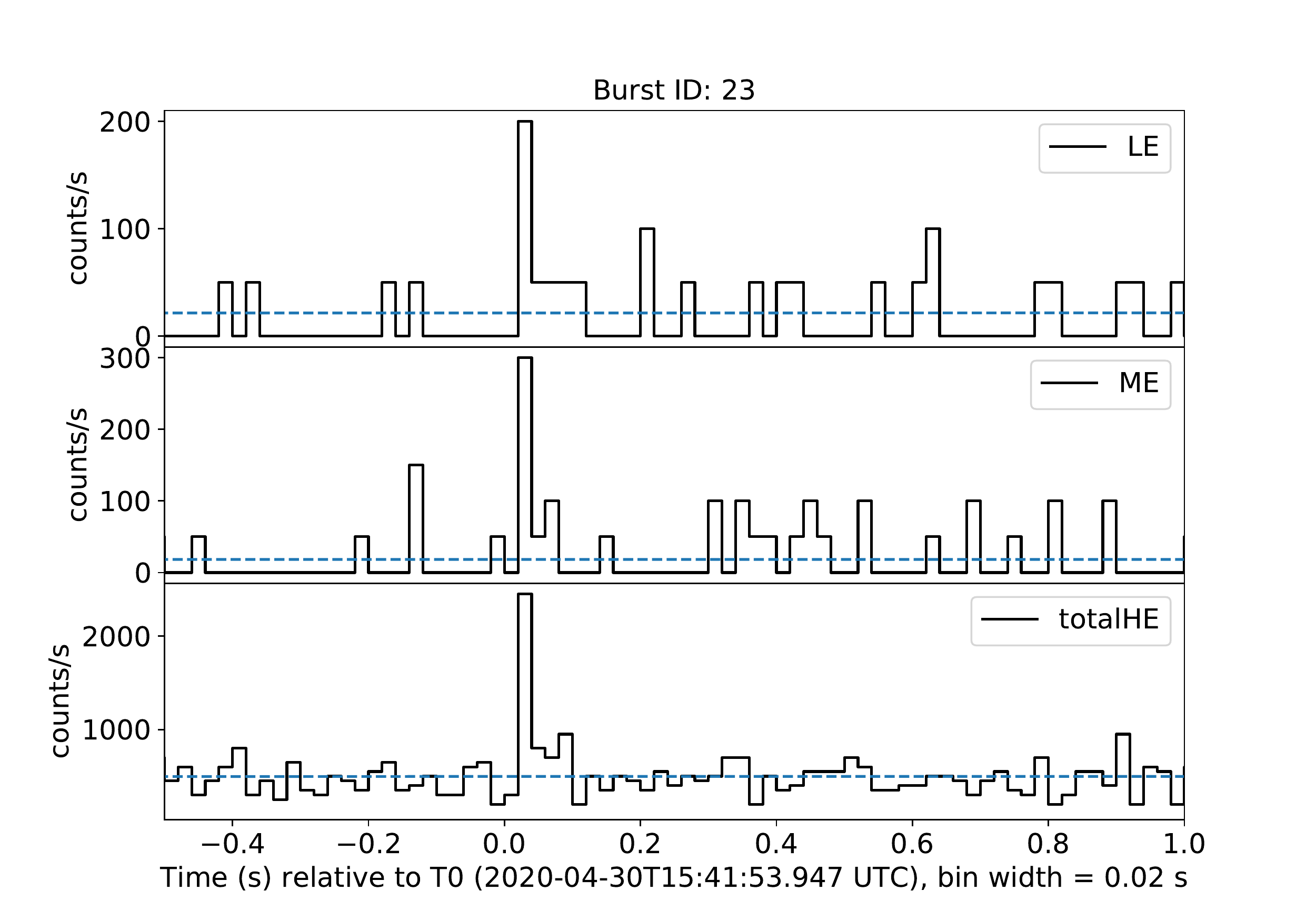}
	\includegraphics[width=60mm]{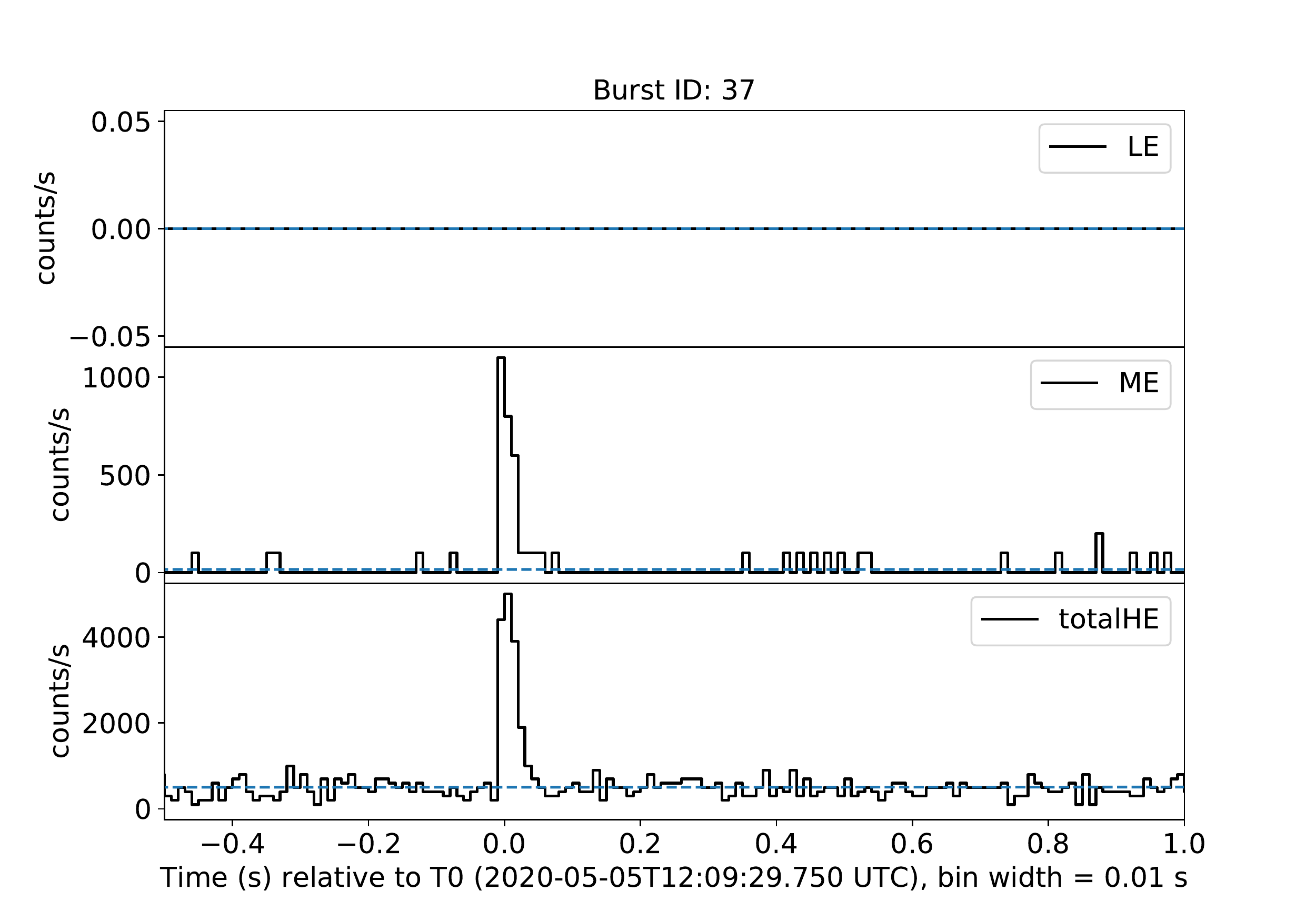}
	
	\vspace{1mm}

	\includegraphics[width=60mm]{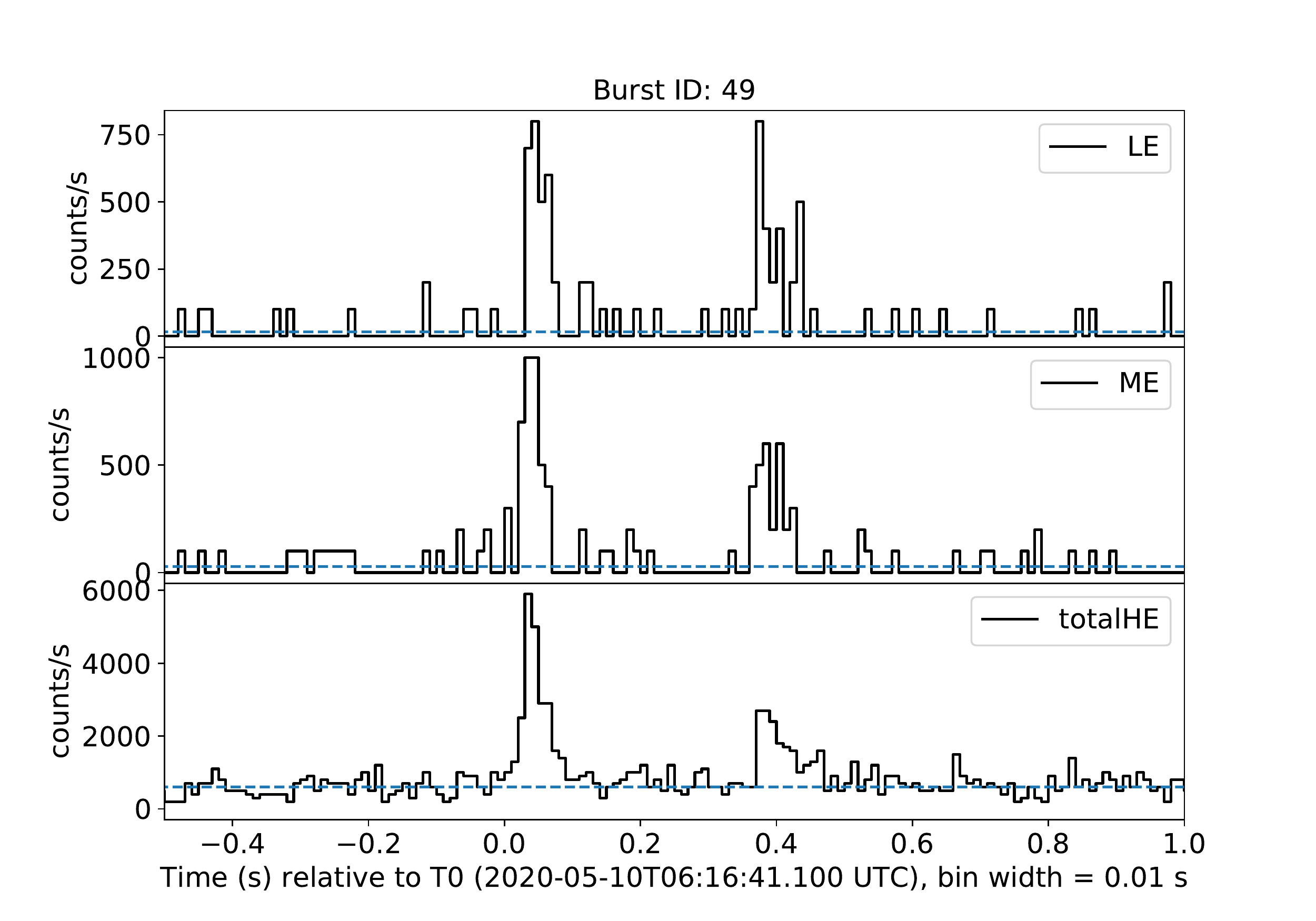}
	\includegraphics[width=60mm]{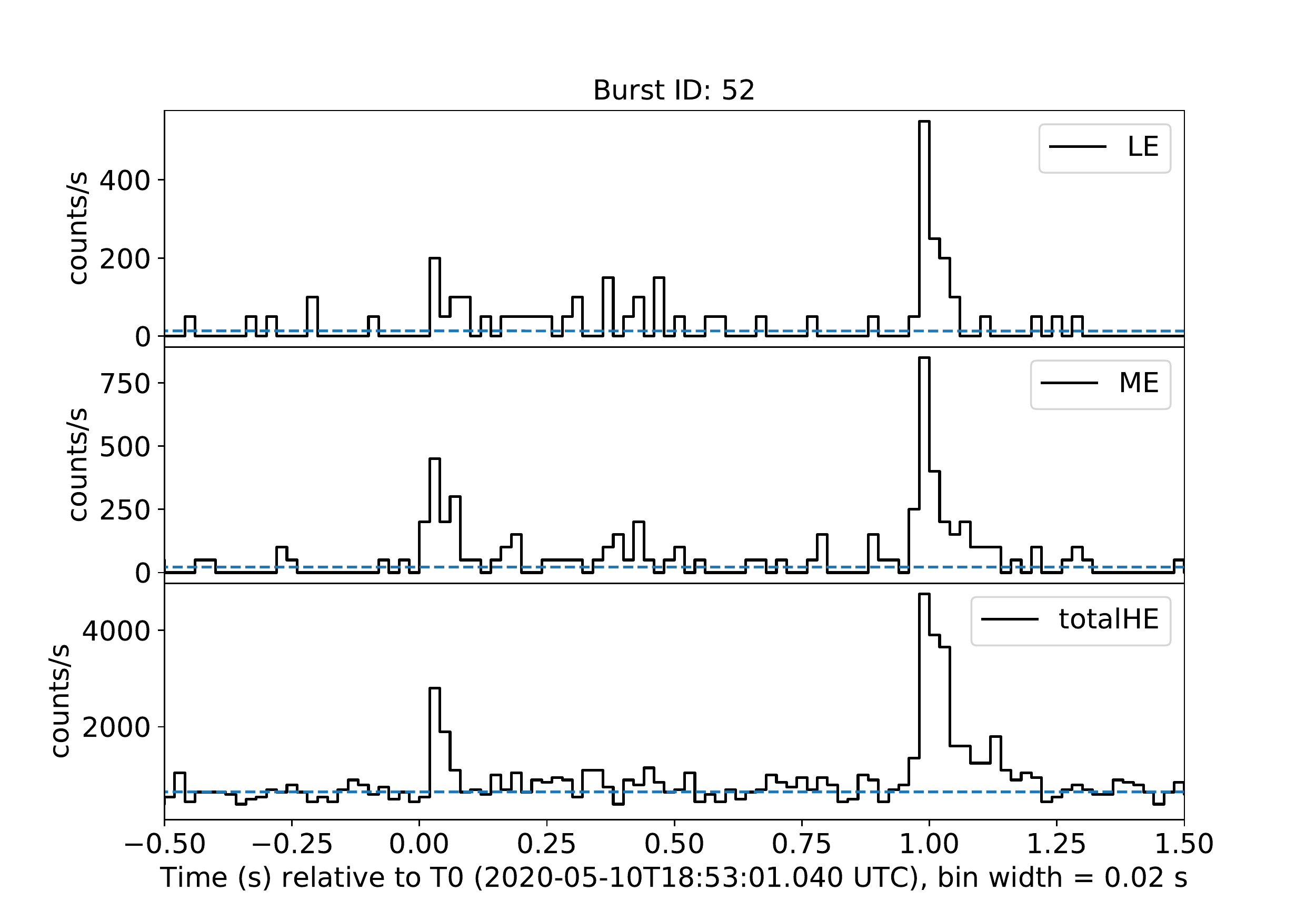}
	\includegraphics[width=60mm]{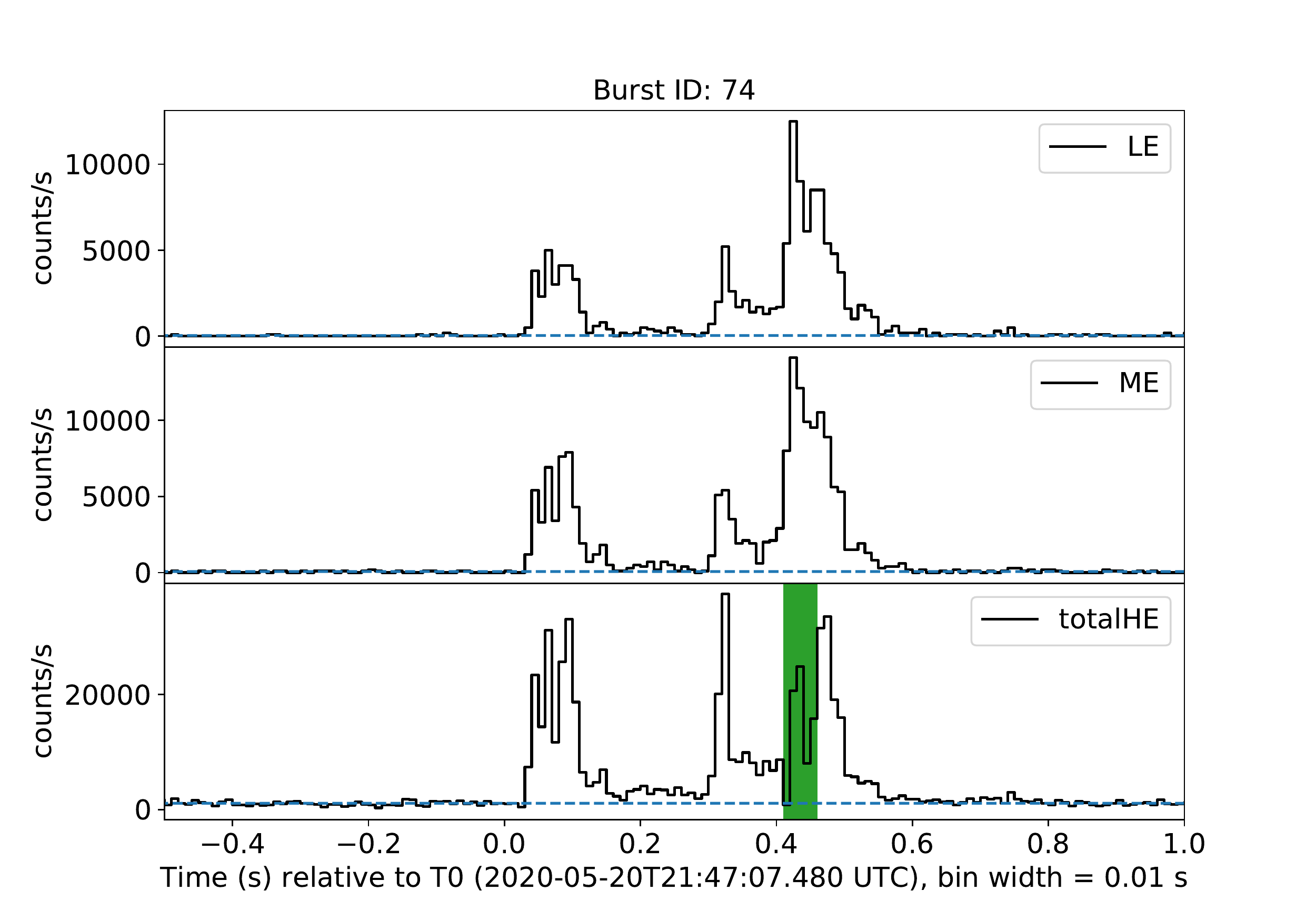}
	
	\vspace{1mm}
	
\caption{\label{fig:Candidates}  Examples of \textit{Insight}-HXMT detected bursts of SGR J1935+2154. For each burst, light curves are shown for LE ( 1$-$10 keV, top panels), ME (10$-$30 keV, middle panels) and HE ( 28$-$250 keV, lower panels). The background of each burst is shown by the blue dotted line.
The events of LE data of burst \#37 are lost due to the instrumental effect (e.g., bright earth, bad events with higher grade). The HE data of burst \#74 suffered from data saturation in the time interval marked as green shadow. Burst \#23 represents weak bursts in our sample with the trigger timescale of 0.01 s and trigger threshold of $\sim$ 5 $\sigma$. }
\end{figure*}
\end{center}

\subsection{Burst Search}  \label{BURST_Search}
The blind search is based on the signal-to-noise ratio (i.e. SNR) method which has been applied to gamma-ray burst search in the HE CsI detectors of \textit{Insight}-HXMT (see \cite{10.1093/mnras/stab2760} for more details). This method is used to search for magnetar bursts in the HE NaI detectors of \textit{Insigth}-HXMT.
In order to unveil the soft and faint bursts that might not be found with only HE detectors (which has higher energy band than that of LE and ME), we use three telescopes (HE, ME and LE) of \textit{Insight}-HXMT to jointly search for bursts (e.g., \citep{Guidorzi_2020,2020Guidorzi}), covering a series of timescales ranging from 5 ms to 40 ms, with two phase offsets (\citep{10.1093/mnras/stab2760}). Given the limited counts for weak bursts, Poisson statistics is assumed (e.g., \citep{2020ApJ...904L..21Y}).

To help to set the search strategies and to estimate the expected counts in LE, ME and HE for substantially weak soft bursts, we simulate the \textit{Insight}-HXMT observations of the SGR J1935+2154 bursts measured by \textit{NICER}. The expected counts are calculated by multiplying the softest spectrum (burst \# 222) from \textit{NICER} (see Table 2 in \cite{2020ApJ...904L..21Y}, i.e. $kT = 0.5_{-0.1}^{+0.1}$ keV, $R^2 = 970_{-560}^{+1300}$ km$^2$) with the responses of LE, ME and HE detectors. It results in 2 counts, 0 counts and 0 counts in LE, ME and HE respectively, with a duration of about 0.2 s, which means 
that there is some signal in LE while no signal in ME and HE for such kind of soft and weak bursts (flux of 10$^{-9}$ erg cm$^{-2}$ s$^{-1}$ in 0.5-10 keV as measured by \textit{NICER}).
However, given the low background of LE (i.e. $\sim 1$ counts in 0.1 s), LE could marginally detect such weak and soft bursts but with a low significance (less than $\sim 3\sigma$).
We note that, the ratio of model predicted counts of LE and ME is roughly to be 1 for bursts with $kT > 2.5$ keV, which constitutes about $10\%$ of the \textit{NICER} samples (see Table 2 in \cite{2020ApJ...904L..21Y}). However, the spectral parameters of \textit{NICER} samples are derived from a relatively limited energy range of 1$-$10 keV.

According to the count ratios between the three telescopes (HE, ME and LE) in the simulations shown above, some bursts with softer spectra are likely detected by LE only. However, the background of LE is very complicated, which is mainly composed of two parts: particle background ( $>$7 keV) and the diffuse X-ray background ( $<$7 keV) \citep{2020JHEAp2724L}, and the background count rate distribution (see Figure \ref{fig:BackgroundDis}) is very wide and far deviating from a simple Poisson distribution. Therefore, it is not easy to identify the SGR J1935+2145 bursts, especially for weak bursts, only using the light curves of LE. We leave the detailed burst search with LE data only to the future work.

Therefore, in this paper, considering the quality of the burst sample and available data, we require that a burst should be observed by at least two telescopes during the burst search and verification, thus three telescopes of \textit{Insight}-HXMT could be divided into four combinations: HE\&ME, HE\&LE, ME\&LE and HE\&ME\&LE. Following the experience in previous search studies \citep{10.1093/mnras/stab2760, Guidorzi_2020,2020Guidorzi}, all un-blinded detector units are summed together for ME and LE, respectively, while the 17 NaI units of HE are used separately.

The detailed search algorithms and trigger criterion are shown in Table \ref{tab:table2}, including different combinations of three telescopes, timescales, phases, the threshold of detector number of HE and net counts of HE, ME and LE. A burst candidate is found when the excess in the net counts of two or more telescopes exceeds a preset threshold of significance. Different polynomial ﬁts are applied to the light curves of each telescope to estimate the background count rate within a 100-s time window centered on the searched time bin. Then the background counts of each time bin are obtained through the interpolation of the background model.

\subsection{Tests with Blank Sky Observations}
The backgrounds of the three telescopes of \textit{Insight}-HXMT are relatively complicated due to their relatively large field of view.
The LE background is mainly caused by the particles in orbit and the diffuse cosmic X-ray background \citep{2020JHEAp2724L}, while that of the ME is primarily contributed by the charged particles \citep{2020JHEAp2744G}.  The background of HE is more complicated than that of LE and ME, since it is dominated by the activated isotopes (with relatively longer decay time) which is related to the passages of SAA, together with other components including particles and albedo scattering \citep{2020JHEAp2714L}.

In order to set a proper search criterion and estimate the significance for each burst candidate, it is important to know how the background of the three telescopes vary with time and the distribution of background counts on those short timescales used in the SGR burst search. To study the background variation, calibrate our search algorithm, and derive the false alarm probability, we execute the same search pipeline to a group of blank sky background observations (i.e. observations of the selected blank sky regions without bright sources), which spans from 2017-11-02T05:00:57.000 to 2018-03-27T05:24:47.000 (a total span time of 909 hours, or 78537.6 ks), as listed in Table \ref{tab:table1}.

According to these blank sky observations, there are significant extra-Poisson components in the background distributions of all three telescopes (see Figure \ref{fig:BackgroundDis}), which are the imprint of the complicated background behaviors. Such complicated distributions of background indeed require one to derive the significance (i.e. false alarm probability) of the searched burst candidate directly from the background data with real variations rather than from the SNR of the searched time bin (\citep{10.1093/mnras/stab2760}).

We calibrate the trigger criterion by searching for any count excess in the blank sky data with the same trigger threshold as used in the SGR burst search. Detailed results are listed in Table \ref{tab:table2}, including the common good time intervals (GTIs) of different telescopes, total numbers of false triggers $N_{\rm FT}$ and false alarm probability ${P_{\rm FA}}$ of the blank sky at four timescales. The false alarm probability (${P_{\rm FA}}$) is defined as,

\begin{equation}
{P_{\rm FA}}= \frac{N_{\rm FT}}{N_{\rm S}}, \label{FAP}
\end{equation}
where $N_{\rm FT}$ and $N_{\rm S}$ represent the numbers of false triggers and the number of searches of different timescales and different combinations of telescopes. The search number is calculated using the length of common good time intervals divide by timescales and taking into account of the phase shifts.

As shown in Table \ref{tab:table2}, based on the detailed calibration with blank sky data, the search threshold for SGR burst is generally equivalent to at least a 5 $\sigma$ detection (Gaussian).

\subsection{Burst Identification}

In the preliminary analysis released on the website of \textit{Insight}-HXMT \footnote{http://hxmtweb.ihep.ac.cn/bursts/392.jhtml}, we reported 133 burst candidates, which triggered the blind search with rough burst identification. In this work, we execute a refined analysis for each burst candidate with the careful removal of various false triggers (including instrumental effects and real bursts from other sources rather than SGR J1935+2154) and a comprehensive processing of all the data of the three telescopes.

Spikes are short and soft pulses in the HE light curve caused by fake events output by the readout electronics when there is a very large energy deposition in HE detectors. The presence of spikes in the HE data is a well-known instrumental effect and has been studied in detail \citep{wu2022removal}. Here, we find that the spike events cannot be removed completely by the current version (version 2.04) of the \textit{HXMTDAS}, which cause some false triggers, as shown in Figure \ref{fig:spike_filter}.

We identify each burst candidate carefully and filter out spike events based on the following features. The spikes usually occur on only one or a few adjacent NaI and CsI detectors (see the right panel in Figure \ref{fig:spike_burst}) with low energy events (i.e. the pulse height less than about channel 35) \citep{wu2022removal}, while the count increases of all non-blinded NaI detectors should be comparable for a SGR J1935+2154 burst (see the left panel in Figure \ref{fig:spike_burst}) since the energy responses are similar for all non-blinded NaI detectors for the targeted source in the pointed observations. This is an important criterion to identify spikes. In addition, there is no spike event on LE and ME data, which can also be used to identify SGR J1935+2154 bursts mostly triggered by HE.

Apart from the SGR bursts, the blind search could unveil various bursts from other sources, including GRBs and TGFs (Terrestrial-Gamma Flashes). The typical duration of short bursts from SGRs ranges from 0.1 s to 1 s \citep{2015ApJS..218...11C,Lin_2020b}.
In contrast, GRBs would appear in a much broader duration ranging from milliseconds to more than thousands of seconds. Indeed, \textit{Insight}-HXMT/HE has been served as a wide-field gamma-ray monitor for GRBs, mainly using the CsI detectors of HE \citep{Zhang_2020,10.1093/mnras/stab2760}. We find that some short GRBs (nominally with duration less than 2 s) would look similar to the SGR bursts in NaI detectors when they incident the HE telescope from the front side. Thus GRBs could be a potential contamination source to the SGR bursts. However, the spectra of GRBs usually extend from tens of keV up to several MeV, much harder than SGR bursts, thus GRBs are mostly detected by CsI instead of NaI and leave much more counts in CsI than NaI detectors. This feature is opposite to the fact that SGR bursts are soft and mostly detected by NaI and barely leave signals in CsI detectors in this pointed observation of SGR J1935+2154. Therefore, we use the light curves of both NaI and CsI detectors to identify and remove GRBs from the burst candidates of SGR J1935+2154.

TGFs are short intense gamma-ray flashes produced in the lightning process in the atmosphere of the Earth \citep{1994STIN...9611316F,2017Natur.551..481E}. Its duration ranges from sub-millisecond to several milliseconds with a hard spectrum in the energy range of hundreds of keV to tens of MeV \citep{1997JGR...102.9659N,2013JGRA..118.3805B}. There are a wealth of TGF detections by \textit{Insight}-HXMT (Yi et al., In Prep.). Similar to the case of GRBs, TGFs are hard in spectrum and leave more counts in CsI than NaI detectors. Moreover, TGFs are much shorter than typical SGR bursts. Based on these characteristics, it is convenient to identify TGFs out of the detected bursts.

In summary, we check each burst candidate found by the blind search and filter out the false triggers including spikes, GRBs and TGFs. We obtained a sample of 75 bursts which are confidently identified as from SGR J1935+2154, to our best knowledge. The burst samples are listed in Table \ref{table4}, including the FRB 200428-Associated burst (i.e. burst \#15).

\begin{figure}
\centering
\begin{tabular}{c}
\includegraphics[width=0.50\textwidth]{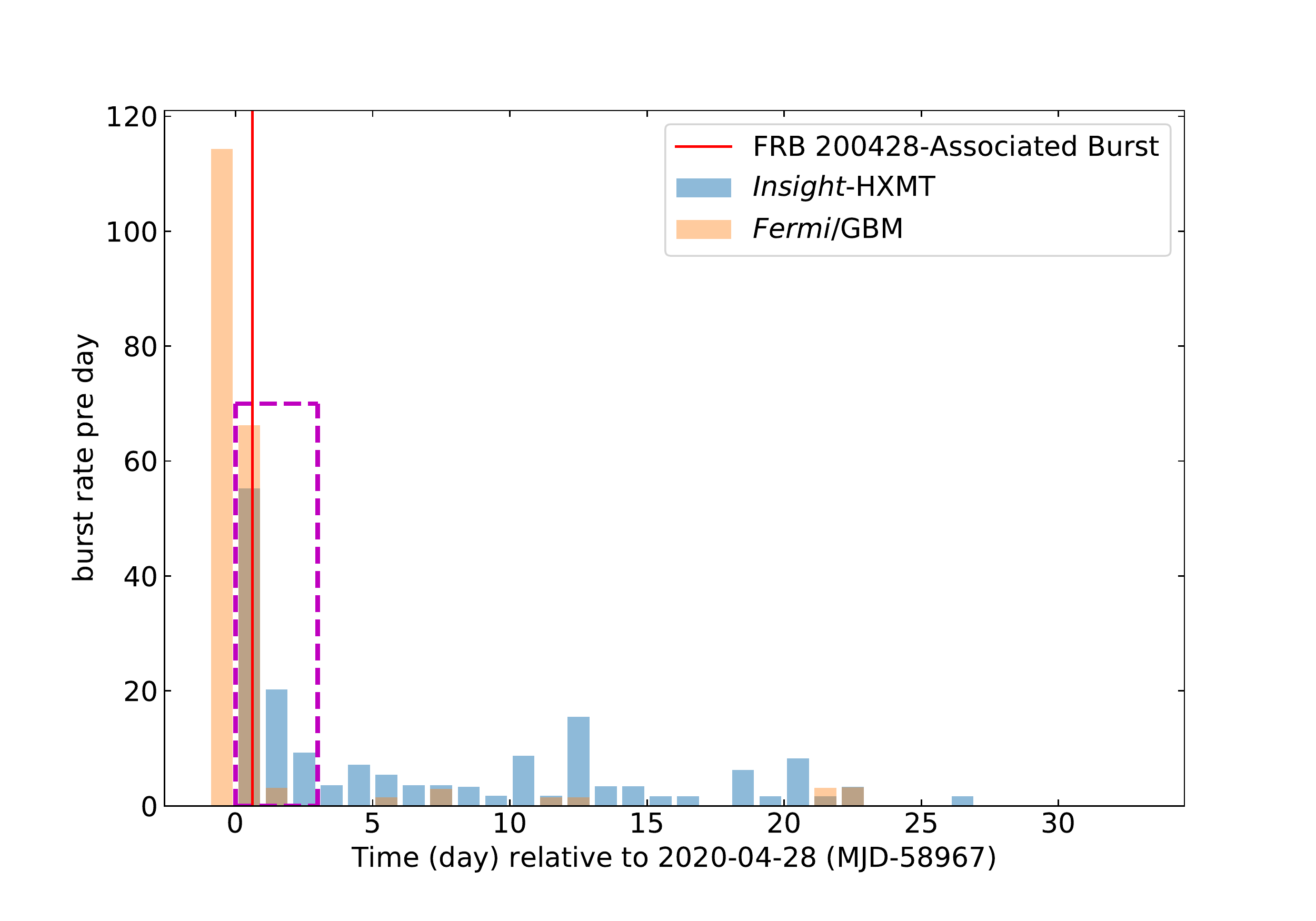}  \\
\end{tabular}
\caption{\label{fig:Num_time} Burst activity evolution of SGR J1935+2154 in 1-day time bin as seen with \textit{Insight}-HXMT and \textit{Fermi}/GBM \citep{Lin_2020b}. The first burst in this activity phase was observed by \textit{Swift}/BAT and \textit{Fermi}/GBM around 2020-04-27T18:26:20.138. The start time of \textit{Insight}-HXMT observation was about 13 hours later. There is no observation from 2020-04-29T12:02:36.000 UTC to 2020-04-30T06:58:23.000 UTC, resulting in incomplete monitoring of SGR J1935+2154 in these three days (purple dashed box) (also see Table \ref{tab:table1}).
The daily burst rate is the number of bursts divided by the effective exposure time when the source is monitored (i.e., the source is not blocked by the Earth and the satellite is not in SAA). The blue and orange bars represent the burst rates of \textit{Insight}-HXMT and \textit{Fermi}/GBM, respectively. The red line is the trigger time of FRB 200428-Associated burst.}
\end{figure}

\section{Catalog Analysis AND RESULTS} \label{sec:result}

As shown above, we identify a total of 75 bursts in the dedicated 33-day ToO observation for SGR J1935+2154. We show light curves of 9 bursts in the sample in Figure \ref{fig:Candidates}. The trigger time of each burst is listed in Table \ref{table4}. Most bursts are detected by all three telescopes, while a few bursts are detected by only one or two telescopes since other telescopes data are unavailable to use (e.g. out of the GTI).

\subsection{Burst Activity}
\label{sec:activity history}
We define the daily burst rate ($R$, in units of bursts per day, 24 hours) as:

\begin{equation}
{R} = \frac{N}{P},
\end{equation}
where $N$ is the observed burst number per day and $P$ represents the percentage of the effective observation time in a full day, excluding the time intervals of Earth blocking of the SGR J1935+2154 and SAA passages of \textit{Insight}-HXMT, during which the instrument (observation) would be turned off.

Note that, due to the schedules of ToO observations, the observation coverage is incomplete in three days (i.e. 2020-04-28, 2020-04-29 and 2020-04-30), for which the burst rate ($R$) also accounts for the non-observation time intervals.
%

As shown in Figure \ref{fig:Num_time}, the burst rate generally decreases with time during the 33-day evolution of burst activity, with a few small re-active peaks. This \textit{Insight}-HXMT dedicated long ToO observation caught up the later part of the most active phase and monitored the entire process from very active to inactive in burst rate of SGR J1935+2154. Thanks to the broad energy band (from 1 keV to 250 keV) and high sensitivity, {Insight}-HXMT detected much more bursts than other instruments (e.g. \textit{Fermi}/GBM), providing a unique valuable burst data set to the community.

\begin{figure}
\centering
\begin{tabular}{c}
\includegraphics[width=0.50\textwidth]{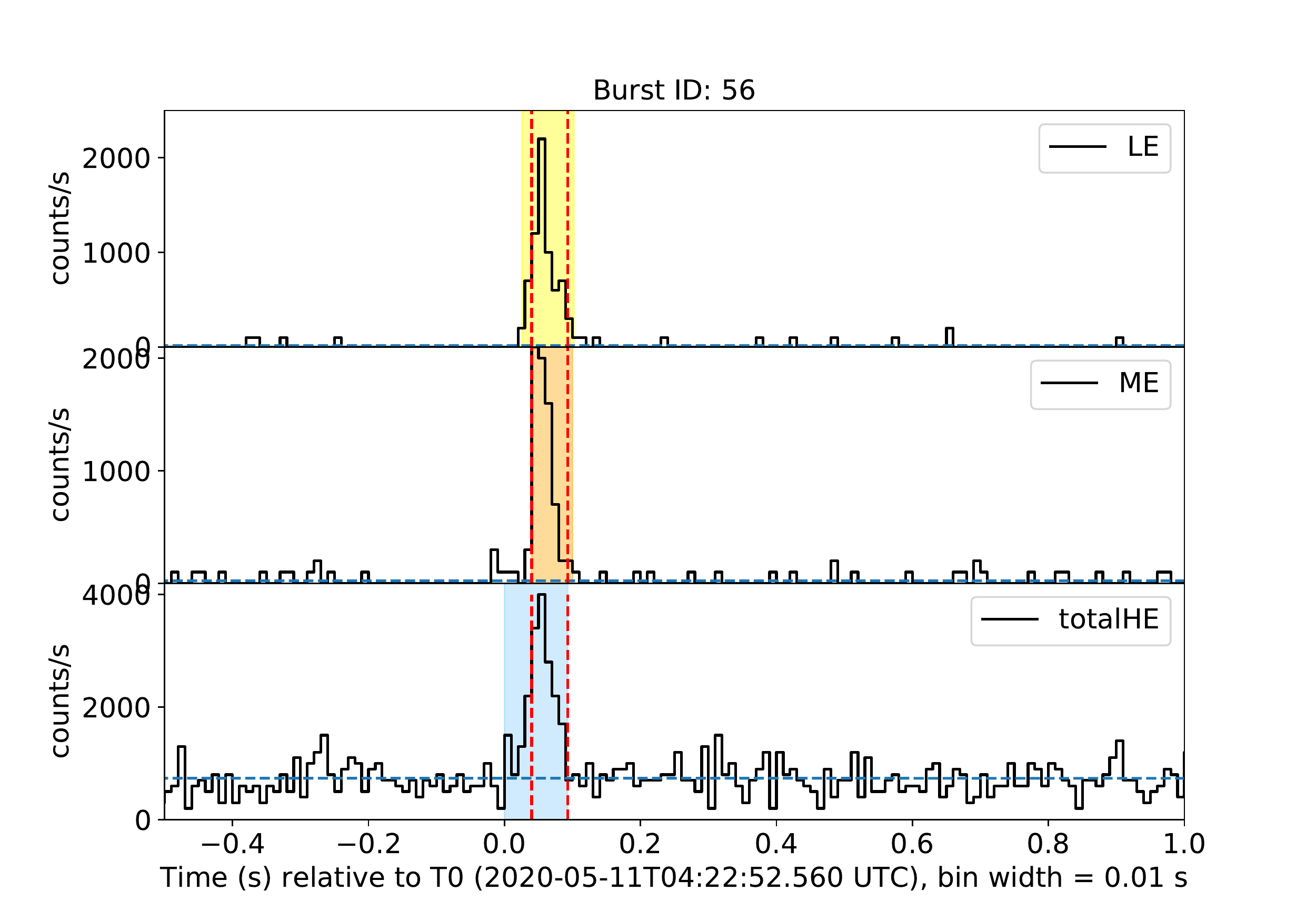}  \\
\end{tabular}
\caption{\label{fig:burst_duration} Illustration of burst duration. Light curves of burst \# 56 are shown in separate panels. The yellow, orange and blue shadows represent the duration of LE, ME and HE, respectively. The red dashed lines indicate the common duration of the three telescopes defined in this work.}
\end{figure}

\begin{figure*}
\centering
\begin{tabular}{cc}
\includegraphics[width=0.50\textwidth]{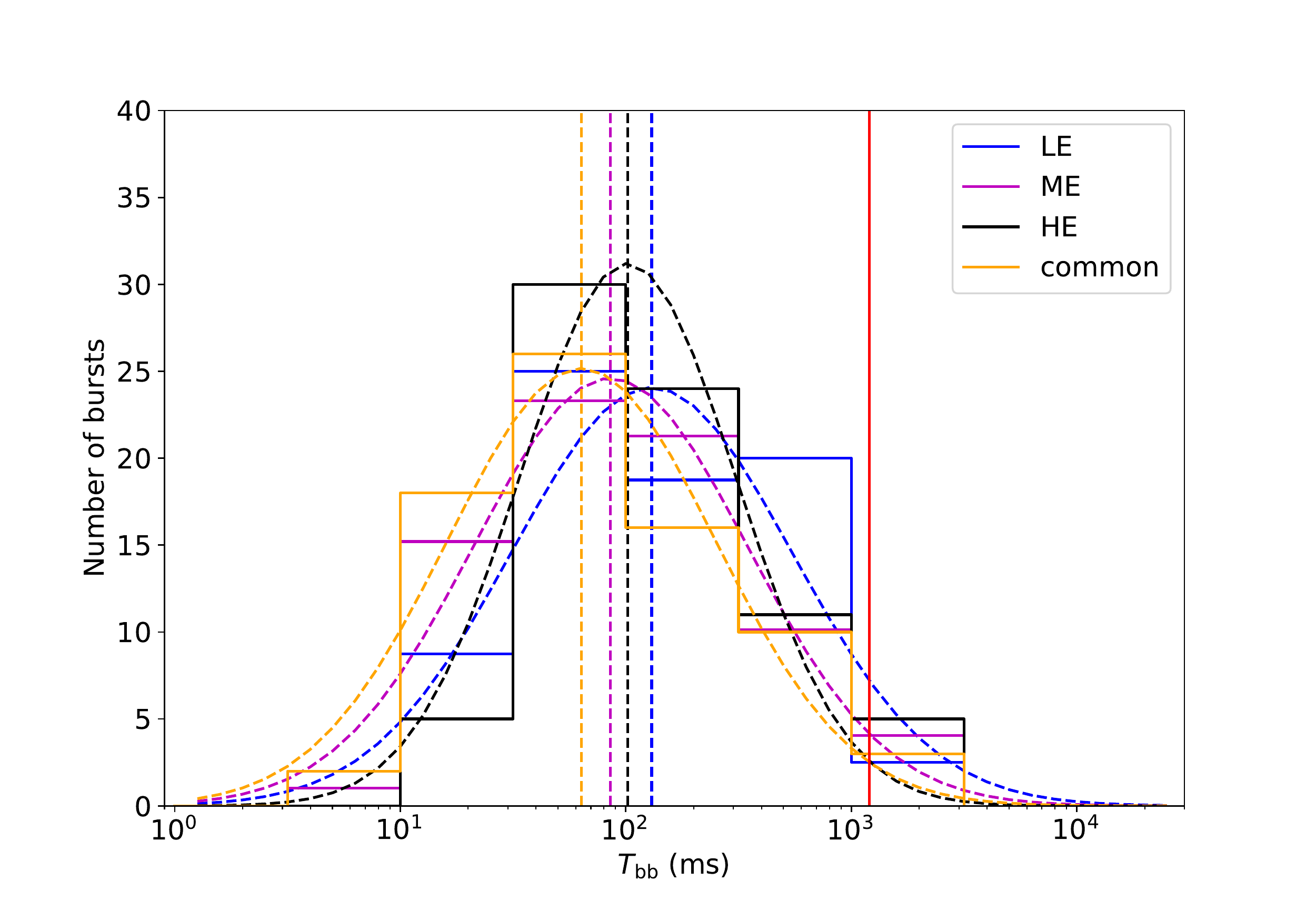} &
\includegraphics[width=0.50\textwidth]{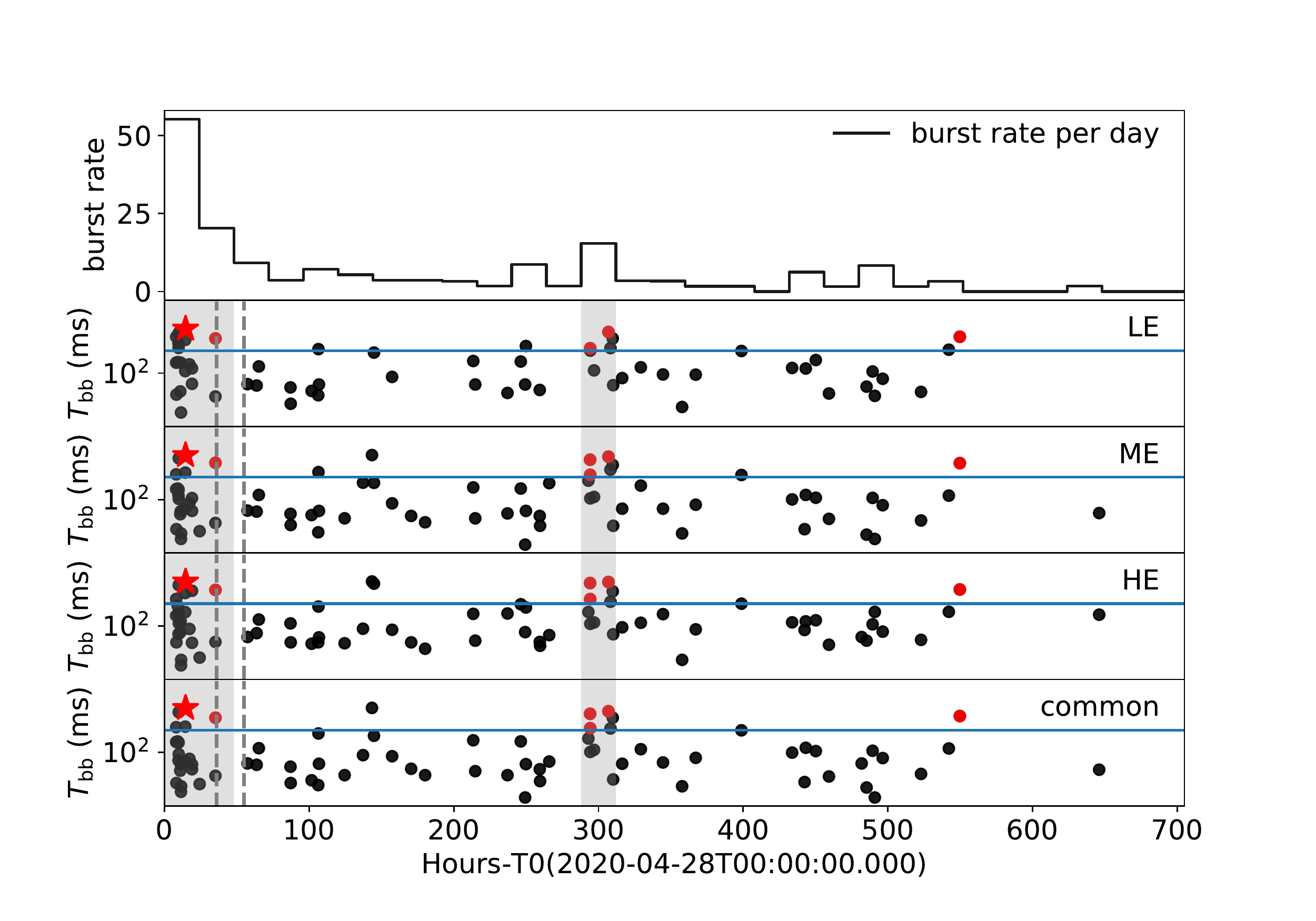}  \\
\end{tabular}
\caption{\label{fig:Duration} \textit{Left}: The burst distribution of the whole burst sample. The best fit log-Gaussian functions and corresponding mean values are dotted lines. The blue, purple, black and orange lines represent the duration of LE, ME, HE and common, respectively. The red line is FRB 200428-Associated Burst.
\textit{Right}: The burst history of SGR J1935+2154 in the top panel.
Other captions of the top panel are the same as Figure \ref{fig:Num_time}.
The scatter of duration versus their trigger time since 2020-04-28T00:00:00 is shown in the four bottom panels. The gray dotted lines represent the time interval of no observation of \textit{Insight}-HXMT (also see Table \ref{tab:table1}).
The red star is the FRB 200428-Associated Burst. Multi-pulse bursts are shown in red. The bursts with longer duration (larger than 350 ms, as marked by horizontal blue lines) are mostly found in the gray shadows.}.
\end{figure*}

\begin{figure}
\centering
\begin{tabular}{c}
\includegraphics[width=0.50\textwidth]{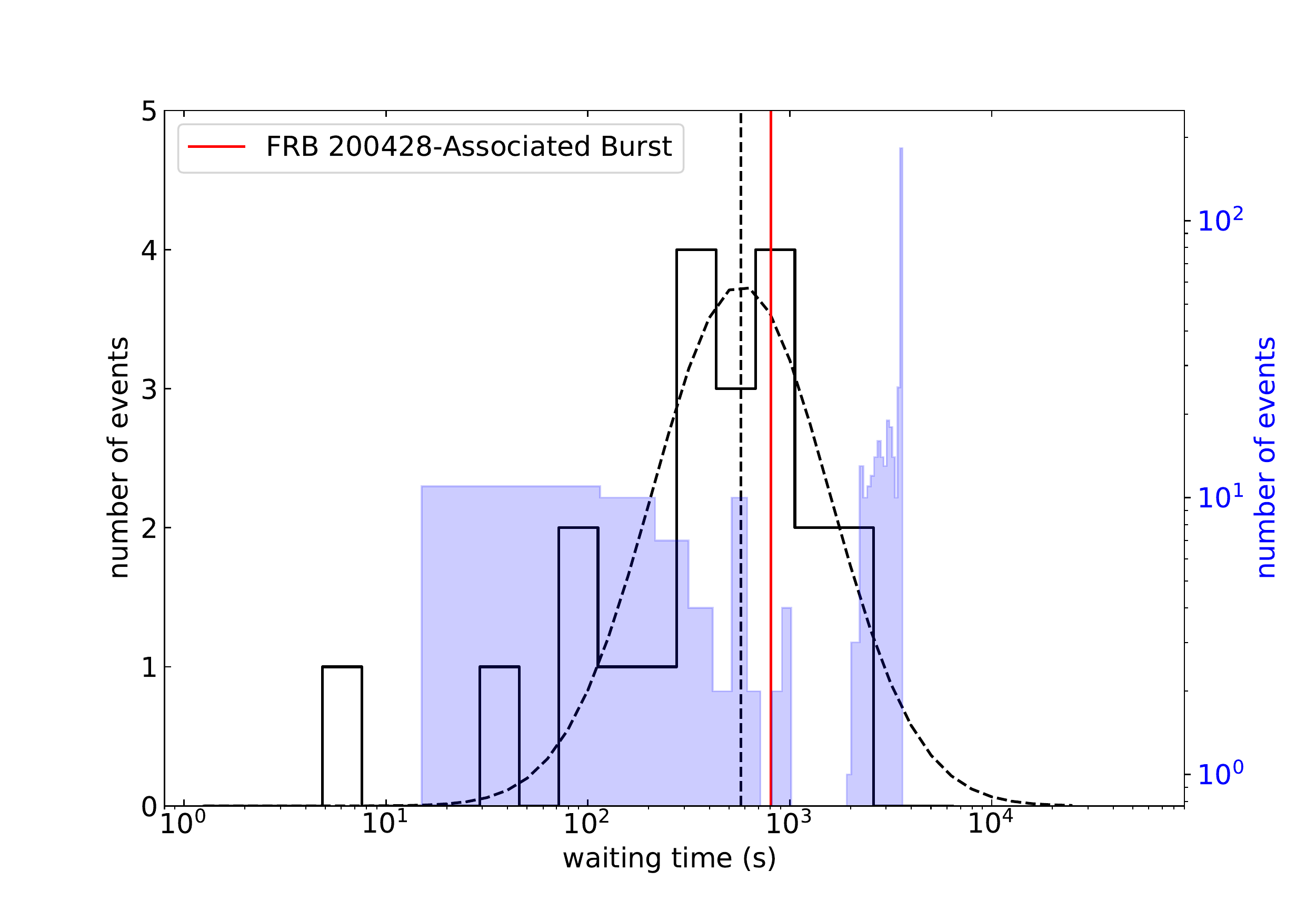} \\
\end{tabular}
\caption{\label{fig:waiting time} The distribution of waiting time for 21 time intervals (i.e., time difference between two successive bursts). The black dashed lines show the best fit log-normal function with $\mu = 572{\pm82}$ s and $\sigma = 0.44{\pm0.06}$
(the standard deviation in the logarithm scale). The distribution of the continuous observation time length is shown in the blue shadows (right vertical axis). The percentage of the continuous observation time length above 1000 s is 87$\%$.
The red line is the FRB 200428-Associated Burst, whose trigger time is used to calculate waiting time for one time interval.}
\end{figure}

\begin{figure*}
\centering
\begin{tabular}{cc}
\includegraphics[width=0.50\textwidth]{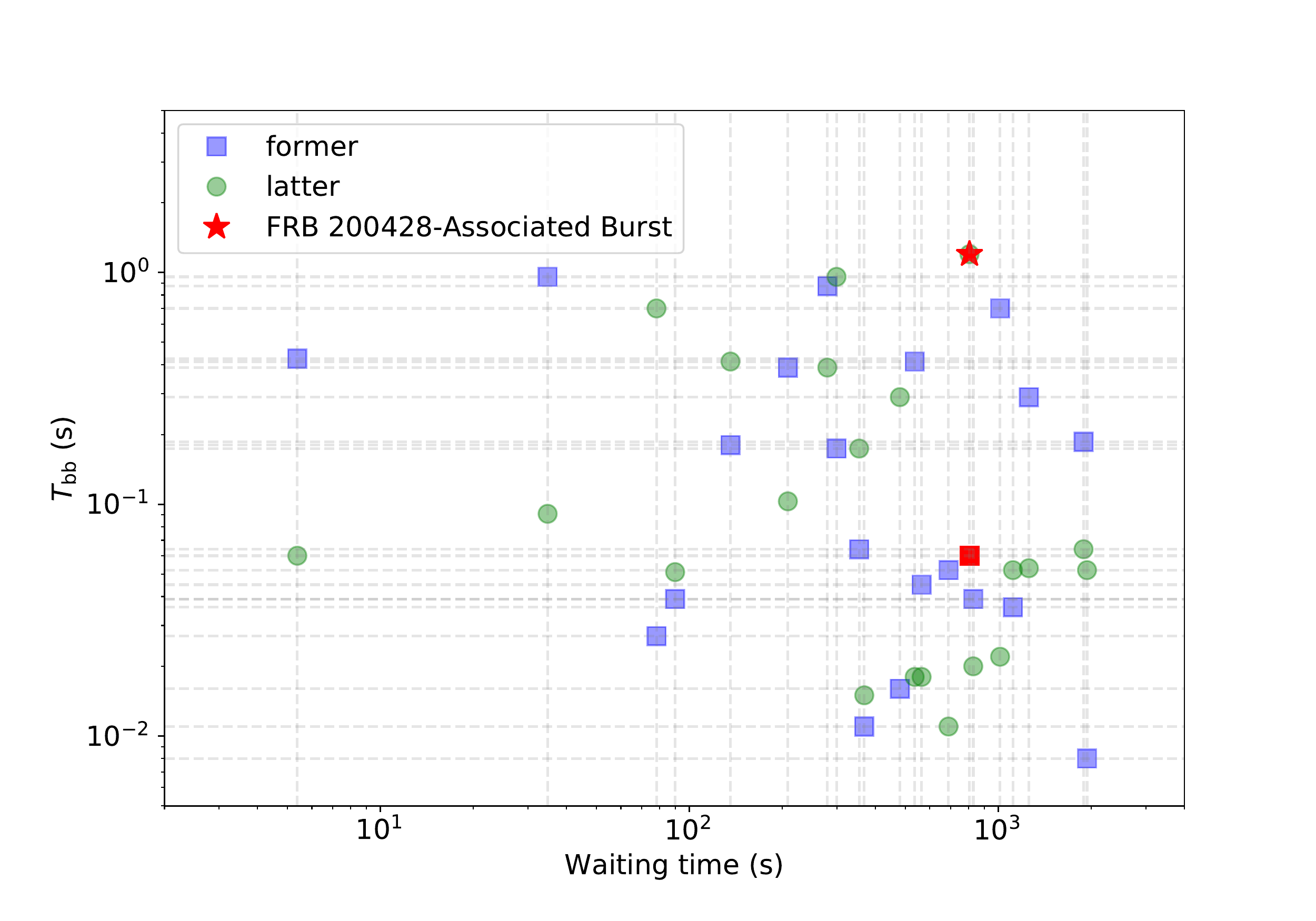} &
\includegraphics[width=0.50\textwidth]{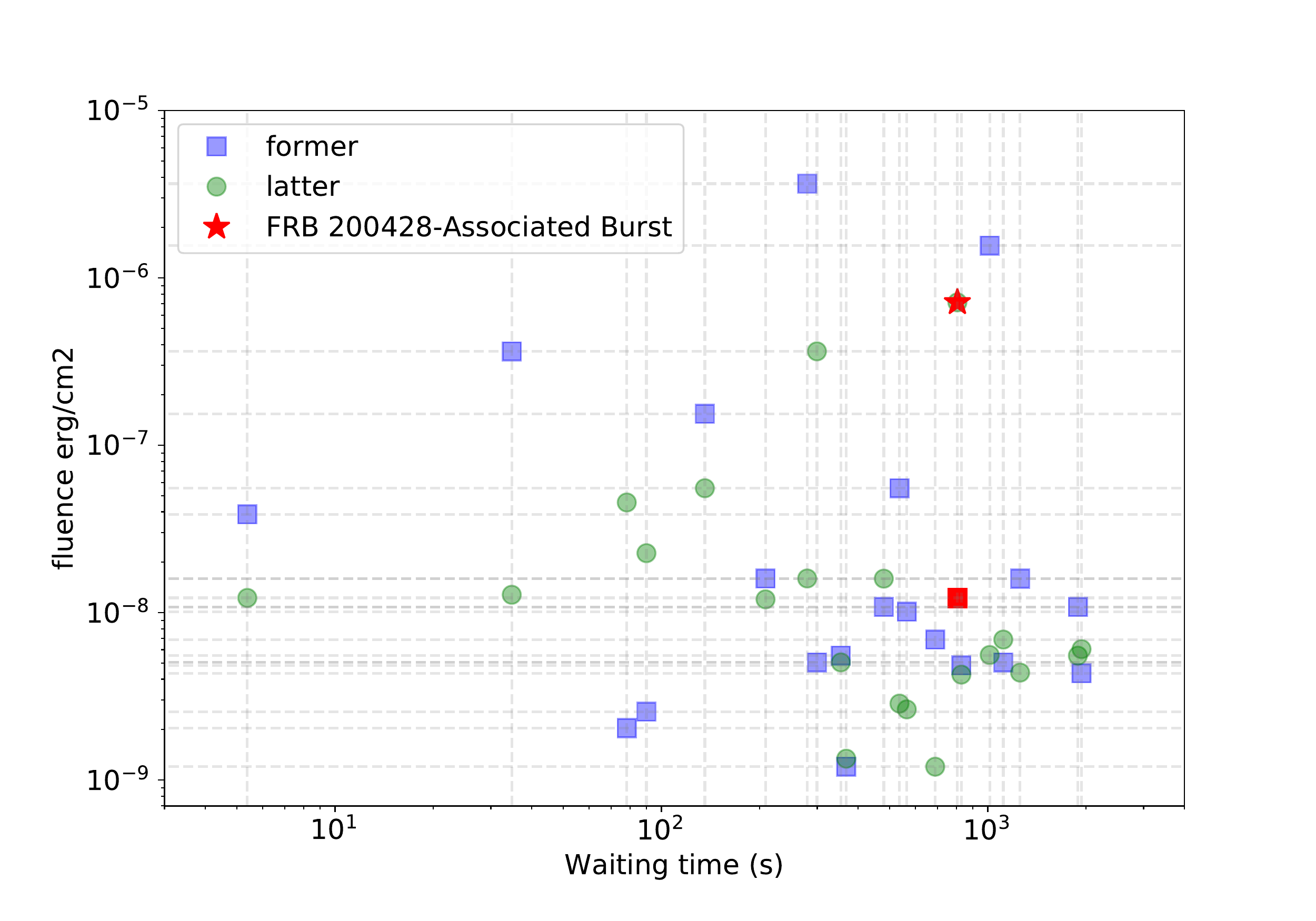}  \\
\end{tabular}
\caption{\label{fig:wt_t_f}\textit{Left}: The scatter plot of the waiting time and common duration ($T_{\rm bb}$) for 21 bursts. \textit{Right}: The scatter plot of the waiting time and fluence (see paper \uppercase\expandafter{\romannumeral2}) for 21 bursts. The blue and green points represent the two bursts used to obtain the waiting time. The red star is the FRB 200428-Associated Burst, which is the latter burst of their pair bursts. The red square is the earlier burst of the FRB 200428-Associated Burst.}
\end{figure*}

\subsection{Burst Duration}
\label{sec:Durations}
Following previous studies \citep{2013Lin}, the bursts duration ${T}_{\rm bb}$ is calculated with the Bayesian blocks method \citep{2013ApJ...764..167S}.
We perform the Bayesian blocks to measure the duration for all bursts using the screened event data of 10-s burst time window, including both pre-burst and post-burst regions. The blocks with a duration longer than 6 s are treated as background, while blocks with duration less than the spin period (i.e., 3.24 s) of SGR J1935+2154 are considered as part of the burst region\citep{2013Lin}. The background count rate is estimated with the mean rate of the background blocks. The burst blocks with less count rate than the background are excluded. Some bursts with multiple pulses (e.g., burst \#49 in Figure \ref{fig:Candidates}) have at least two subsequent blocks, along with bursts-free (or quiescent) intervals, so the burst duration is defined as the time length from the start of the first burst block to the end of the last block within the burst time window, which represents the entire burst duration. As an example of the definition of burst duration, light curves of burst \# 56 are shown in Figure \ref{fig:burst_duration}.

The duration  (${T}_{\rm bb}$) of each burst measured by the three telescopes of \textit{Insight}-HXMT are listed in Tabel \ref{table4}. The burst duration distributions of LE, ME and HE are shown in the left panel of Figure \ref{fig:Duration}. These distributions are well fit with lognormal functions with the mean value of $130{\pm25}$ ms and $\sigma = 0.62{\pm0.08}$ for LE, $86{\pm7}$ ms and $\sigma = 0.61{\pm0.03}$ for ME, $102{\pm11}$ ms and $\sigma = 0.48{\pm0.04}$ for HE.
\footnote{All errors presented in this paper are for 1 $\sigma$ confidence level, unless otherwise stated.} We define the common time interval (duration) of LE, ME and HE, where the start time is the maximum start time while the end time is the minimum end time of three telescopes. The common duration is also listed in Table \ref{table4}. For bursts detected by less than three telescopes, the common duration is obtained with the same manner. We also plot the distribution of common duration, which can be described with a Gaussian function of $\mu = 64{\pm8}$ ms and $\sigma = 0.59{\pm0.05}$, as shown in the left panel of Figure \ref{fig:Duration}.

We also checked the evolution of the burst duration, as shown in the right panel of Figure \ref{fig:Duration}. We find that longer bursts (with duration larger than 350 ms) mostly occurred during those time intervals with higher burst rate (i.e., the gray shadows).

\begin{table*}[htbp]
\centering
\caption{Burst duration measured by \textit{Insight}-HXMT telescopes.}
\label{duration_table}
\begin{tabular}{ p{2cm}<{\centering}|  p{2.4cm}<{\centering}   p{1.8cm} <{\centering} p{1.8cm}<{\centering} |p{2.4cm}<{\centering}   p{1.8cm} <{\centering} p{1.8cm}<{\centering} } 
\toprule
\multicolumn{1}{c}{Samples} & \multicolumn{3}{c}{Individual Sample$^1$} & 
\multicolumn{3}{c}{Common Sample$^2$}  \\
\hline
telescope &  number of bursts &  $\mu$ (ms) &  $\sigma^3$ &  number of bursts &  $\mu$ (ms) &  $\sigma^3$ \\  
LE &  60  & $130{\pm25}$  & $0.62{\pm0.08}$ &  60  & $130{\pm25}$  & $0.62{\pm0.08}$\\
ME  & 74 & $86{\pm7}$  & $0.61{\pm0.03}$ &  60  & $92{\pm6}$  & $0.59{\pm0.02}$\\
HE & 75 & $102{\pm11}$  & $0.48{\pm0.04}$ &  60  & $119{\pm13}$  & $0.48{\pm0.04}$\\
Common$^4$  & 75 & $64{\pm8}$  & $0.59{\pm0.05}$ &  60  & $78{\pm6}$  & $0.61{\pm0.03}$\\
\hline
\end{tabular}
\tablecomments{$^1$ Individual burst samples detected by LE, ME and HE respectively.\\
$^2$ Common burst sample which is jointly detected by all three telescopes.\\
$^3$ The standard deviation in the logarithm scale.\\
$^4$ Common duration of the three telescopes.\\}
\end{table*}

\begin{figure*}
\centering
\begin{tabular}{cc}
\includegraphics[width=0.50\textwidth]{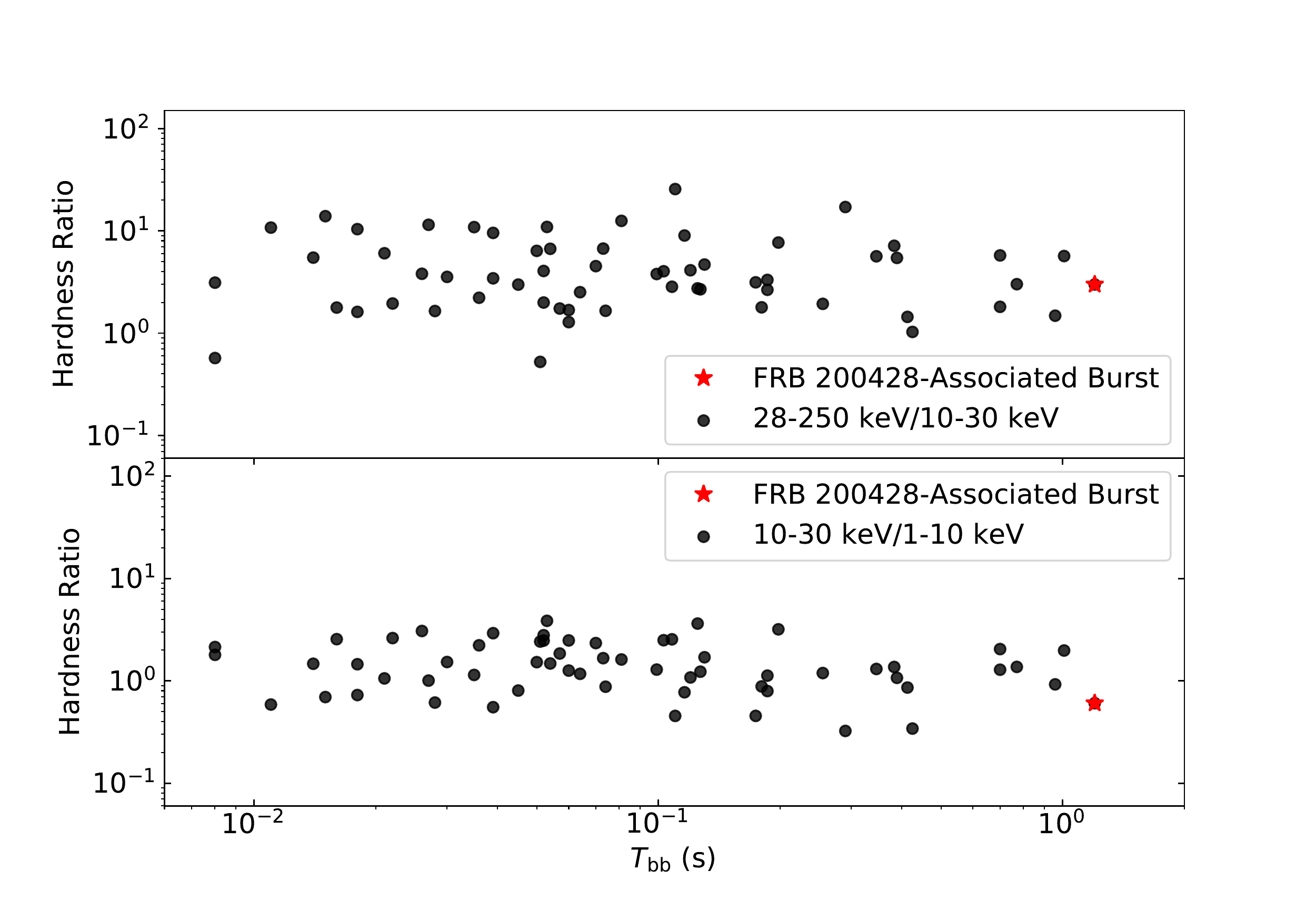} &
\includegraphics[width=0.50\textwidth]{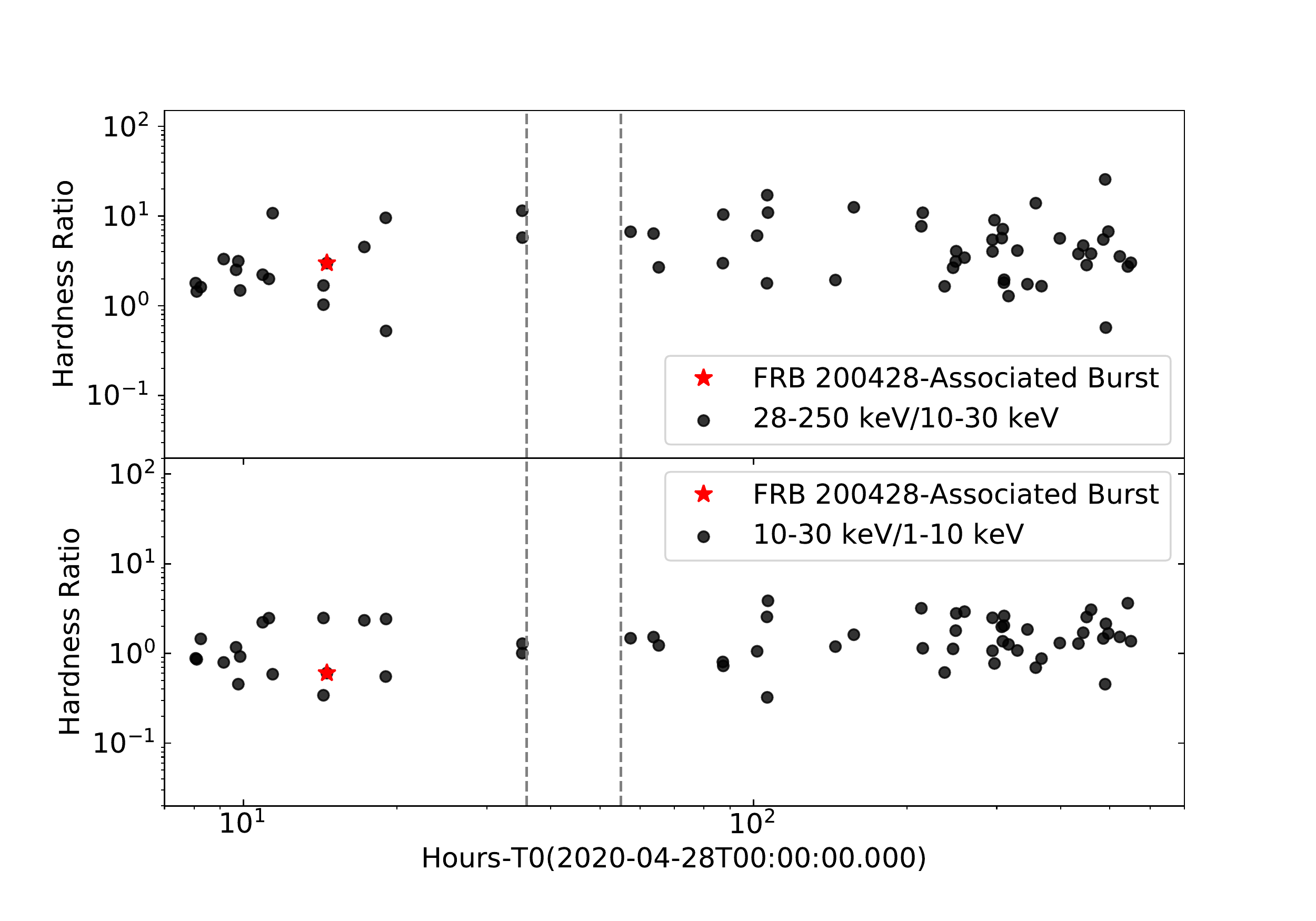}  \\
\end{tabular}
\caption{\label{fig:HR} \textit{Left}: The scatter plot of hardness ratio versus duration. \textit{Right}: The evolution of hardness ratio of the each burst in 28$-$250 keV and 10$-$30 keV, 10$-$30 keV and 1$-$10 keV.
The gray dotted lines represent the time interval of no observation (also see Table \ref{tab:table1}). The red star is the FRB 200428-Associated Burst.}
\end{figure*}

\subsection{Burst Waiting Time}
\label{sec:Waiting Time}

We measure the waiting time (${\delta}_{t}$) between successive bursts which fall within an uninterrupted GTI,
\begin{equation}
{\delta}_{t} = {t}_{i+1} - {t}_{i},
\end{equation}
where ${t}_{i+1}$ and ${t}_{i}$ represent the trigger times of ($i+1$)-th (latter) and $i$-th (previous) bursts, respectively.

To obtain the continuous observation time intervals (i.e. without interruption) with \textit{Insight}-HXMT, we exclude the time intervals when the satellite passes through SAA or SGR J1935+2154 is blocked by the Earth. During those continuous observation time intervals, there are 21 waiting times ${\delta}_{t}$.
The distribution of these waiting times is shown in Figure \ref{fig:waiting time}, which ranges from 5.38 s to 1935.94 s and could be fitted to a log-normal function with a peak of $572{\pm82}$ s and $\sigma = 0.44{\pm0.06}$. The length distribution of the continuous observation time interval is also shown in Figure \ref{fig:waiting time}, which naturally provides the maximum value of the waiting time.

The scatter plot of waiting time versus $T_{\rm bb}$ and fluence (see paper \uppercase\expandafter{\romannumeral2}) is shown in Figure \ref{fig:wt_t_f}. We find that there is no significant correlation between the waiting time and the fluence or the burst duration either for the previous burst or the latter burst.

\subsection{Burst Hardness Ratio}
\label{sec:HR}

The net count of each burst is estimated as:
\begin{equation}
{C}_{i} = {S}_{i} - {B}_{i},
\end{equation}
where ${S}_{i}$ and ${B}_{i}$ represent the total counts and background counts in the burst duration, respectively. The net counts are computed for HE, ME, and LE, respectively.

The hardness ratio is the ratio of the net count rates in different energy bands.
We derive the hardness ratio between 10$-$30 keV to 1$-$10 keV with ME and LE data, and hardness ratio between 28$-$250 keV to 10$-$30 keV with ME and HE data. There are 60 bursts detected by all three telescopes of \textit{Insight}-HXMT, which could be used for hardness ratio study.
The relationship between common burst duration versus hardness ratio of these 60 bursts is shown in the left panel of Figure \ref{fig:HR}. Hardness ratios between 10$-$30 keV to 1$-$10 keV and between 28$-$250 keV to 10$-$30 keV of these bursts spread over a large range from 0.52 to 25.60 and from 0.32 to 3.86, respectively.
There is no correlation between the duration and the hardness ratio.
As shown in the right panel of Figure \ref{fig:HR}, there is also no significant trend in the evolution of the burst hardness across the observation time.


Note that there are 7 bursts (see Table \ref{table4}) for which HE or LE data suffered from data saturation \citep{2020JHEAp..26...58X}. Thus, the hardness ratio calculation is not applicable for them. We report the detailed spectral analysis to study the hardness of the bursts in paper \uppercase\expandafter{\romannumeral2}, considering the data saturation and deadtime effects.

\section{DISCUSSIONS} \label{sec:discussion}
In the following, we compare our results to SGR J1935+2154 bursts observed by other instruments (e.g. GBM, \textit{NICER}) and other magnetars.

\textit{Insight}-HXMT started this dedicated 33-day ToO observation of SGR J1935+2154 about 13 hours after the initial \textit{Fermi}/GBM and \textit{Swift}/BAT triggers, and detected a total of 75 bursts with the burst daily rate (i.e. number of bursts divided by the effective exposure time per day) varying from $\sim$ 56 bursts day$^{-1}$ to none burst, as shown in Figure \ref{fig:Num_time}.
We note that during this observation of SGR J1935+2154 with \textit{Insight}-HXMT, only 12 bursts were detected by \textit{Fermi}/GBM \citep{Lin_2020b}. There are 7 bursts jointly observed by \textit{Insight}-HXMT and \textit{Fermi}/GBM. The remaining 5 bursts are invisible to \textit{Insight}-HXMT (e.g., the satellite flies through SAA, or SGR J1935+2154 is blocked by the earth).
The much higher burst rate of \textit{Insight}-HXMT than \textit{Fermi}/GBM is consistent with the fact that \textit{Insight}-HXMT has a much wider energy band coverage (1 - 250 keV), was pointed to the source, and thus has much higher sensitivity than \textit{Fermi}/GBM. Although there is no burst jointly observed by \textit{Insight}-HXMT and \textit{NICER}, we find that \textit{Insight}-HXMT could detect most of the bursts found by \textit{NICER} \citep{2020ApJ...904L..21Y}, based on the simulations of \textit{Insight}-HXMT observations (see Section \ref{BURST_Search}).

Regarding the burst morphology of the light curve structure, all bursts could be generally classified to two groups: single-pulse bursts or multi-pulse bursts. We require that multi-pulse bursts must have distinctive pulses separated by non-emission time intervals (i.e. quiescent period).
We find that in this sample the majority of bursts (70/75, $\sim$ 93$\%$) shows single-pulse (e.g., burst \#2, burst \#4 in Figure \ref{fig:Candidates}), and there are only 5 multi-pulse bursts, most of which are shown in Figure \ref{fig:Candidates} (burst \#21, burst \#49, burst \#52, burst \#74). We notice that the spectra (see paper \uppercase\expandafter{\romannumeral2}) of the multi-pulse bursts are similar to those of the single-pulse bursts.

The burst duration follows a log-normal distribution, as the case for SGR J1935+2154 bursts seen by other instruments as well as the bursts from other magnetars (e.g., \citep{2015ApJS..218...11C, 2020ApJ...904L..21Y}.
There are 7 bursts which are detected by both \textit{Insight}-HXMT and GBM. The burst duration of \textit{Insight}-HXMT is slightly longer than that of GBM for these 7 bursts \citep{Lin_2020b}. Note that the three telescopes of \textit{Insight}-HXMT have narrow fields of view and low background instruments working in 1 to 250 keV, while \textit{Fermi}/GBM is an all-sky monitor with relatively high background operating above 8 keV. This is likely responsible for the difference of duration between  \textit{Insight}-HXMT and \textit{Fermi}/GBM. We also find that the duration of LE is slightly longer on average than ME and HE (see Table \ref{duration_table} for details), which is also evident in the statistical properties of the whole burst sample. Nevertheless, some bursts show interesting behavior that the HE duration is longer than that of LE and ME (e.g. burst \#52 in Figure \ref{fig:Candidates}). The different duration measured by LE, ME and HE is likely a reflection of the spectral evolution. Interestingly, we notice 
that the durations are different for the bursts in the periods with different burst rates;
the burst with a longer duration and with multi-pulse tends to occur in the more active periods (see Figure \ref{fig:Duration}).

The distribution of waiting time between successive SGR J1935+2154 bursts observed by \textit{Insight}-HXMT is well described by a log-Gaussian function (Figure \ref{fig:waiting time}), which is also found in other observations of SGR J1935+2154 and that of other magnetars (e.g. \cite{1999Ersin,2000180620,2020MNRAS.491.1498C}). \cite{2020ApJ...904L..21Y} reported the waiting time distribution of SGR J1935+2154 bursts during burst storms.
\cite{1999Ersin} showed that the waiting time of SGR 1900+14 is a log-normal function with a peak of 49 s. \cite{2000180620} showed the waiting time of SGR 1806-20 is also a log-normal function that peaks at 97 s.
The peak waiting time of SGR J1935+2154 observed by \textit{Insight}-HXMT is 571 s, much longer than those measurements mentioned above, which mainly reflects that the SGR J1935+2154 burst rate (burst activity) is relatively low during this ToO observation. In principle, the waiting time between bursts depends on the burst activity, instrumental sensitivity and the time length of continuous observation (see Figure \ref{fig:waiting time}).

There are three reported radio bursts from SGR J1935+2154. One was detected by Five-hundred-metre Aperture Spherical radio telescope (FAST) \citep{2018IMMag..19..112L} at 2020-04-30T21:43:00.500 UTC \citep{fastatel2020b}, and the other two were detected by Westerbork RT1 (Wb) at 2020-05-24T22:19:19.674 UT and 2020-05-24T22:19:21.070 UT, respectively \citep{2020Kirsten}. At the time of the first radio burst of FAST, SGR J1935+2154 was visible to \textit{Insight}-HXMT, which operated as normal, while it was invisible to \textit{Insight}-HXMT for the latter two radio bursts. We check our burst list and do not find any X-ray bursts associated with these radio bursts.
We estimate the 3 $\sigma$ upper limit flux is 2.5 $\times 10^{-9}$ erg cm$^{-2}$ s$^{-1}$ in the 1$-$250 keV energy band, assuming the same spectral parameters with that of the FRB 200428-Associated burst \citep{2021NatAs...5..378L} and 1 s duration timescale.

\section{Summary} \label{sec:summary}
In this paper, we analyze the \textit{Insight}-HXMT data of the dedicated 33-day ToO monitoring of SGR J1935+2154, initiated by the breakthrough discovery of the X-ray burst associated with the first FRB confirmed from a magnetar. Based on the preliminary analysis, we refine the burst search and classification, and verify a total of 75 bursts using the available data of all three telescopes (HE, ME and LE) of \textit{Insight}-HXMT. Thanks to the broad energy band and high sensitivity, \textit{Insight}-HXMT detects a burst sample significant larger than that of other instruments during the same observation period.

We implemented extensive studies on the statistical characteristics of these 75 bursts, including the burst duration, waiting time and hardness ratio. The duration follows log-normal distribution with mean values of 130.22 ms, 85.52 ms and 101.95 ms for LE, ME and HE, respectively. The waiting time distribution is well fitted with a log-normal function with the mean value of 571.53 s. Moreover, the two hardness ratios spread extensively from 0.52 to 25.60 and from 0.32 to 3.86, respectively.
We find that most bursts are single-pulsed with only 5 with multiple pulses. Those bursts with a longer duration tend to happen when the magnetar experiences burst active periods. So far, we do not find any X-ray bursts associated with any reported radio bursts except for FRB 200428.

Given that there are many multi-wavelength observations to SGR J1935+2154 during April and May of 2020, this \textit{Insight}-HXMT dedicated 33-day long ToO observation provides an unprecedented data set in 1-250 keV for more detailed analysis to depict the full picture of the burst activity evolution of SGR J1935+2154 for both pre- and post- the generation of FRB.

\begin{longrotatetable}
\centering
\begin{deluxetable*}{lcccccccccccc}
\tablecaption{The \textit{Insight}-HXMT SGR J1935+2154 burst list of the dedicated 33-day ToO observation from April 28 to June 1, 2020. \label{table4}}
\tablewidth{700pt}
\tabletypesize{\scriptsize}
\tablehead{
\colhead{ID} & \colhead{Trigger time} & 
\colhead{${T}^1_{\rm bb}$} & \colhead{${T}^2_{\rm {st}}$} & 
\colhead{${T}^3_{\rm bb(HE)}$} &  \colhead{${T}^4_{\rm{st}(HE)}$} & \colhead{${T}^3_{\rm bb(ME)}$} &  \colhead{${T}^4_{\rm {st}(ME)}$} &
\colhead{${T}^3_{\rm bb(LE)}$ } & \colhead{${T}^4_{\rm {st}(LE)}$} &
\colhead{$C^5_{\rm HE}$} &  \colhead{$C^5_{\rm ME}$} & \colhead{$C^5_{\rm LE}$}
\\ 
\colhead{} & \colhead{in UTC} & \colhead{(s)} & \colhead{(s)} & 
\colhead{(s)} & \colhead{(s)} & \colhead{(s)} &
\colhead{(s)} & \colhead{(s)} & \colhead{(s)} & \colhead{counts} & \colhead{counts} & \colhead{counts} }
\startdata
1$^{S1}$	&	2020-04-28T08:03:34.300	&	0.180 	&	0.010 	&	0.180 	&	0.010 	&	0.180 	&	0.010 	&	0.180 	&	0.010 	&	603 	&	337 	&	382 	\\
2 	&	2020-04-28T08:05:50.080	&	0.413 	&	0.026 	&	0.455 	&	0.024 	&	0.413 	&	0.026 	&	0.768 	&	-0.010 	&	431 	&	271 	&	587 	\\
3 	&	2020-04-28T08:14:45.985	&	0.018 	&	0.020 	&	0.040 	&	0.011 	&	0.019 	&	0.019 	&	0.030 	&	0.020 	&	36 	&	10 	&	11 	\\
4 	&	2020-04-28T09:08:44.280	&	0.186 	&	0.036 	&	0.308 	&	0.034 	&	0.186 	&	0.036 	&	0.186 	&	0.036 	&	224 	&	41 	&	51 	\\
5 	&	2020-04-28T09:40:10.980	&	0.064 	&	0.004 	&	0.064 	&	0.004 	&	0.127 	&	0.004 	&	0.531 	&	-0.416 	&	40 	&	31 	&	112 	\\
6 	&	2020-04-28T09:46:05.300	&	0.174 	&	-0.137 	&	0.202 	&	-0.142 	&	0.174 	&	-0.137 	&	0.412 	&	-0.178 	&	58 	&	16 	&	84 	\\
7$^{S2G}$	&	2020-04-28T09:51:04.634	&	0.958 	&	-0.096 	&	0.993 	&	-0.131 	&	1.014 	&	-0.096 	&	0.958 	&	-0.096 	&	2512 	&	1731 	&	1770 	\\
8$^*$	&	2020-04-28T09:51:39.394	&	0.091 	&	0.013 	&	0.117 	&	0.013 	&	0.104 	&	0.000 	&	\nodata 	&	\nodata 	&	149 	&	34 	&	\nodata 	\\
9 	&	2020-04-28T10:54:23.850	&	0.036 	&	0.036 	&	0.070 	&	0.033 	&	0.044 	&	0.033 	&	0.036 	&	0.036 	&	117 	&	33 	&	12 	\\
10 	&	2020-04-28T11:12:58.520	&	0.052 	&	0.036 	&	0.130 	&	0.020 	&	0.052 	&	0.036 	&	0.180 	&	0.010 	&	120 	&	24 	&	34 	\\
11 	&	2020-04-28T11:24:28.120	&	0.011 	&	0.014 	&	0.011 	&	0.014 	&	0.011 	&	0.014 	&	0.011 	&	0.014 	&	19 	&	2 	&	3 	\\
12$^*$	&	2020-04-28T11:30:36.180	&	0.015 	&	0.012 	&	0.015 	&	0.012 	&	0.015 	&	0.012 	&	\nodata 	&	\nodata	&	44 	&	2 	&	\nodata 	\\
13 	&	2020-04-28T14:20:52.519	&	0.425 	&	-0.003 	&	0.637 	&	-0.003 	&	0.455 	&	-0.033 	&	0.645 	&	-0.036 	&	182 	&	126 	&	524 	\\
14 	&	2020-04-28T14:20:57.900	&	0.060 	&	0.047 	&	0.218 	&	0.042 	&	0.060 	&	0.047 	&	0.112 	&	0.044 	&	285 	&	47 	&	35 	\\
15$^{S2}$ 	&	2020-04-28T14:34:24.150	&	1.20 	&	-0.20 	&	1.20 	&	-0.20 	&	1.20 	&	-0.20 	&	1.20 	&	-0.20 	&	6805 	&	2275 	&	3762 	\\
16 	&	2020-04-28T17:15:26.237	&	0.070 	&	0.015 	&	0.085 	&	0.002 	&	0.083 	&	0.002 	&	0.164 	&	0.015 	&	149 	&	32 	&	27 	\\
17 	&	2020-04-28T19:00:29.948	&	0.039 	&	0.017 	&	0.039 	&	0.017 	&	0.109 	&	-0.025 	&	0.055 	&	0.005 	&	38 	&	11 	&	10 	\\
18 	&	2020-04-28T19:01:59.850	&	0.051 	&	0.041 	&	0.721 	&	-0.500 	&	0.053 	&	0.039 	&	0.131 	&	0.041 	&	585 	&	82 	&	84 	\\
19$^*$	&	2020-04-29T00:17:40.942	&	0.017 	&	0.029 	&	0.017 	&	0.029 	&	0.017 	&	0.029 	&	\nodata 	&	\nodata 	&	50 	&	7 	& \nodata	\\
20 	&	2020-04-29T11:12:39.397	&	0.027 	&	0.016 	&	0.041 	&	0.008 	&	0.027 	&	0.016 	&	0.027 	&	0.016 	&	58 	&	3 	&	3 	\\
21$^G$	&	2020-04-29T11:13:57.650	&	0.700 	&	-0.480 	&	0.758 	&	-0.485 	&	0.788 	&	-0.499 	&	0.700 	&	-0.480 	&	1001 	&	181 	&	125 	\\
22 	&	2020-04-30T09:25:22.750	&	0.054 	&	0.005 	&	0.054 	&	0.005 	&	0.054 	&	0.005 	&	0.054 	&	0.005 	&	33 	&	5 	&	3 	\\
23 	&	2020-04-30T15:41:53.947	&	0.050 	&	0.023 	&	0.067 	&	0.023 	&	0.051 	&	0.023 	&	0.050 	&	0.023 	&	51 	&	6 	&	4 	\\
24 	&	2020-04-30T17:12:52.837	&	0.127 	&	0.012 	&	0.144 	&	-0.002 	&	0.130 	&	0.009 	&	0.146 	&	0.012 	&	367 	&	123 	&	113 	\\
25 	&	2020-05-01T15:05:56.635	&	0.045 	&	0.016 	&	0.116 	&	0.016 	&	0.045 	&	0.016 	&	0.045 	&	0.016 	&	170 	&	22 	&	28 	\\
26 	&	2020-05-01T15:15:20.876	&	0.018 	&	0.026 	&	0.040 	&	0.010 	&	0.024 	&	0.024 	&	0.018 	&	0.026 	&	97 	&	6 	&	6 	\\
27 	&	2020-05-02T05:40:53.151	&	0.021 	&	0.013 	&	0.037 	&	0.006 	&	0.042 	&	-0.008 	&	0.037 	&	0.013 	&	42 	&	8 	&	7 	\\
28 	&	2020-05-02T10:17:26.000	&	0.016 	&	0.035 	&	0.040 	&	0.033 	&	0.016 	&	0.035 	&	0.029 	&	0.035 	&	167 	&	38 	&	27 	\\
29 	&	2020-05-02T10:25:25.777	&	0.290 	&	0.019 	&	0.298 	&	0.011 	&	0.467 	&	0.017 	&	0.385 	&	0.019 	&	269 	&	25 	&	63 	\\
30 	&	2020-05-02T10:46:20.850	&	0.053 	&	-0.006 	&	0.053 	&	-0.006 	&	0.053 	&	-0.006 	&	0.053 	&	-0.006 	&	51 	&	5 	&	1 	\\
31$^*$	&	2020-05-03T04:30:59.050	&	0.028 	&	0.008 	&	0.038 	&	0.008 	&	0.035 	&	0.001 	&	\nodata 	&	\nodata	&	67 	&	3 	&	\nodata 	\\
32$^*$	&	2020-05-03T17:12:55.600	&	0.086 	&	-0.041 	&	0.086 	&	-0.041 	&	0.258 	&	-0.041 	&	\nodata 	&	\nodata 	&	72 	&	19 	&	\nodata 	\\
33$^{GS1*}$	&	2020-05-03T23:25:13.250	&	1.216 	&	0.000 	&	1.216 	&	0.000 	&	1.216 	&	0.000 	&	\nodata 	&	\nodata 	&	5129 	&	2593 	&	\nodata 	\\
34 	&	2020-05-04T00:48:07.343	&	0.255 	&	0.028 	&	1.076 	&	0.026 	&	0.256 	&	0.027 	&	0.317 	&	0.028 	&	406 	&	50 	&	52 	\\
35 	&	2020-05-04T13:20:00.700	&	0.081 	&	-0.027 	&	0.081 	&	-0.027 	&	0.081 	&	-0.027 	&	0.081 	&	-0.027 	&	82 	&	7 	&	4 	\\
36$^*$	&	2020-05-05T02:30:28.450	&	0.040 	&	0.017 	&	0.040 	&	0.017 	&	0.040 	&	0.017 	&	\nodata 	&	\nodata 	&	39 	&	3 	&	\nodata 	\\
37$^*$	&	2020-05-05T12:09:29.750	&	0.028 	&	-0.006 	&	0.028 	&	-0.006 	&	0.028 	&	-0.006 	&	\nodata 	&	\nodata 	&	93 	&	15 	&	\nodata	\\
38 	&	2020-05-06T21:25:16.350	&	0.198 	&	-0.127 	&	0.198 	&	-0.127 	&	0.198 	&	-0.127 	&	0.198 	&	-0.127 	&	189 	&	25 	&	8 	\\
39 	&	2020-05-06T22:48:21.550	&	0.035 	&	0.016 	&	0.044 	&	0.007 	&	0.035 	&	0.016 	&	0.053 	&	0.016 	&	55 	&	4 	&	5 	\\
40 	&	2020-05-07T21:05:41.345	&	0.028 	&	0.025 	&	0.202 	&	0.010 	&	0.046 	&	0.021 	&	0.033 	&	0.020 	&	245 	&	34 	&	40 	\\
41 	&	2020-05-08T06:17:16.589	&	0.186 	&	0.026 	&	0.339 	&	-0.099 	&	0.186 	&	0.026 	&	0.192 	&	0.020 	&	602 	&	124 	&	114 	\\
42 	&	2020-05-08T09:17:05.185	&	0.008 	&	0.023 	&	0.071 	&	-0.022 	&	0.008 	&	0.023 	&	0.053 	&	-0.004 	&	77 	&	3 	&	10 	\\
43 	&	2020-05-08T09:49:21.134	&	0.052 	&	0.012 	&	0.283 	&	0.006 	&	0.053 	&	0.011 	&	0.458 	&	0.012 	&	158 	&	7 	&	23 	\\
44 	&	2020-05-08T19:23:36.028	&	0.039 	&	0.031 	&	0.041 	&	0.031 	&	0.040 	&	0.030 	&	0.039 	&	0.031 	&	57 	&	16 	&	5 	\\
45$^*$	&	2020-05-08T19:37:25.270	&	0.020 	&	0.011 	&	0.033 	&	0.011 	&	0.023 	&	0.008 	&	\nodata 	&	\nodata 	&	53 	&	13 	&	\nodata 	\\
46$^*$	&	2020-05-09T01:56:38.750	&	0.060 	&	0.000 	&	0.060 	&	0.000 	&	0.252 	&	-0.164 	&	\nodata 	&	\nodata 	&	118 	&	13 	&	\nodata 	\\
47$^*$	&	2020-05-10T05:00:28.195	&	0.218 	&	-0.040 	&	0.219 	&	-0.040 	&	0.289 	&	-0.111 	&	\nodata 	&	\nodata 	&	340 	&	77 	&	\nodata	\\
48$^{S1*}$	&	2020-05-10T06:12:01.622	&	0.873 	&	0.021 	&	1.113 	&	-0.219 	&	0.936 	&	0.021 	&	\nodata 	&	\nodata 	&	12208 	&	7265 	&	\nodata 	\\
49 	&	2020-05-10T06:16:41.100	&	0.389 	&	0.034 	&	0.452 	&	0.018 	&	0.401 	&	0.022 	&	0.404 	&	0.034 	&	337 	&	55 	&	52 	\\
50 	&	2020-05-10T06:20:09.400	&	0.103 	&	0.019 	&	0.113 	&	0.014 	&	0.107 	&	0.015 	&	0.354 	&	0.019 	&	199 	&	47 	&	62 	\\
51 	&	2020-05-10T08:55:46.300	&	0.116 	&	0.031 	&	0.122 	&	0.027 	&	0.116 	&	0.031 	&	0.117 	&	0.031 	&	317 	&	34 	&	44 	\\
52 	&	2020-05-10T18:53:01.040	&	1.008 	&	0.036 	&	1.180 	&	0.010 	&	1.110 	&	0.000 	&	1.008 	&	0.036 	&	507 	&	84 	&	39 	\\
53 	&	2020-05-10T20:16:22.000	&	0.383 	&	0.137 	&	0.390 	&	0.130 	&	0.537 	&	0.134 	&	0.408 	&	0.137 	&	168 	&	32 	&	18 	\\
54$^{S2G}$	&	2020-05-10T21:51:16.221	&	0.700 	&	0.030 	&	0.700 	&	0.030 	&	0.700 	&	0.030 	&	0.700 	&	0.030 	&	8550 	&	4726 	&	2312 	\\
55 	&	2020-05-10T22:08:09.000	&	0.022 	&	0.026 	&	0.063 	&	0.025 	&	0.023 	&	0.025 	&	0.051 	&	0.026 	&	77 	&	15 	&	12 	\\
56 	&	2020-05-11T04:22:52.560	&	0.053 	&	0.040 	&	0.093 	&	0.000 	&	0.060 	&	0.040 	&	0.076 	&	0.026 	&	132 	&	67 	&	67 	\\
57 	&	2020-05-11T17:15:43.320	&	0.120 	&	0.018 	&	0.120 	&	0.018 	&	0.219 	&	0.015 	&	0.139 	&	0.018 	&	231 	&	102 	&	60 	\\
58 	&	2020-05-12T08:35:19.700	&	0.057 	&	0.030 	&	0.195 	&	-0.097 	&	0.060 	&	0.027 	&	0.093 	&	0.030 	&	229 	&	41 	&	34 	\\
59 	&	2020-05-12T21:47:43.340	&	0.015 	&	0.002 	&	0.015 	&	0.002 	&	0.015 	&	0.002 	&	0.015 	&	0.002 	&	28 	&	2 	&	3 	\\
60 	&	2020-05-13T07:12:57.543	&	0.074 	&	0.018 	&	0.083 	&	0.009 	&	0.075 	&	0.018 	&	0.092 	&	0.013 	&	56 	&	30 	&	43 	\\
61 	&	2020-05-14T14:49:22.000	&	0.346 	&	0.030 	&	0.350 	&	0.030 	&	0.397 	&	0.029 	&	0.347 	&	0.029 	&	1242 	&	250 	&	167 	\\
62 	&	2020-05-16T01:50:23.542	&	0.099 	&	0.008 	&	0.123 	&	0.008 	&	0.101 	&	0.006 	&	0.134 	&	-0.014 	&	292 	&	63 	&	65 	\\
63$^*$	&	2020-05-16T10:26:32.309	&	0.019 	&	0.031 	&	0.080 	&	0.020 	&	0.019 	&	0.031 	&	\nodata 	&	\nodata 	&	108 	&	5 	&	\nodata 	\\
64 	&	2020-05-16T11:16:17.000	&	0.130 	&	0.030 	&	0.130 	&	0.030 	&	0.130 	&	0.026 	&	0.130 	&	0.026 	&	425 	&	91 	&	53 	\\
65 	&	2020-05-16T18:12:52.080	&	0.108 	&	0.033 	&	0.138 	&	0.033 	&	0.111 	&	0.030 	&	0.209 	&	0.031 	&	674 	&	191 	&	141 	\\
66 	&	2020-05-17T03:18:10.320	&	0.026 	&	0.037 	&	0.035 	&	0.033 	&	0.034 	&	0.029 	&	0.032 	&	0.037 	&	59 	&	15 	&	5 	\\
67$^+$	&	2020-05-18T01:54:21.550	&	0.054 	&	0.000 	&	0.054 	&	0.000 	&	\nodata 	&	\nodata 	&	\nodata 	&	\nodata 	&	59 	&	\nodata 	&	\nodata 	\\
68 	&	2020-05-18T05:17:57.715	&	0.014 	&	0.024 	&	0.044 	&	0.020 	&	0.014 	&	0.024 	&	0.047 	&	0.024 	&	101 	&	6 	&	13 	\\
69 	&	2020-05-18T09:27:59.151	&	0.110 	&	-0.001 	&	0.110 	&	-0.001 	&	0.110 	&	-0.001 	&	0.110 	&	-0.001 	&	50 	&	2 	&	4 	\\
70 	&	2020-05-18T11:00:41.150	&	0.008 	&	0.032 	&	0.221 	&	-0.155 	&	0.011 	&	0.029 	&	0.028 	&	0.032 	&	77 	&	7 	&	8 	\\
71 	&	2020-05-18T16:28:18.300	&	0.073 	&	0.012 	&	0.073 	&	0.012 	&	0.073 	&	0.012 	&	0.073 	&	0.012 	&	71 	&	11 	&	6 	\\
72$^G$	&	2020-05-19T18:57:36.300	&	0.030 	&	0.020 	&	0.046 	&	0.016 	&	0.031 	&	0.019 	&	0.035 	&	0.020 	&	359 	&	68 	&	50 	\\
73$^G$	&	2020-05-20T14:10:49.780	&	0.125 	&	0.038 	&	0.222 	&	-0.004 	&	0.125 	&	0.038 	&	0.373 	&	-0.010 	&	1013 	&	208 	&	171 	\\
74$^{S1G}$	&	2020-05-20T21:47:07.480	&	0.770 	&	0.040 	&	0.775 	&	0.035 	&	0.773 	&	0.038 	&	0.770 	&	0.040 	&	5060 	&	1676 	&	1219 	\\
75$^*$	&	2020-05-24T22:05:03.480	&	0.038 	&	0.002 	&	0.189 	&	0.002 	&	0.047 	&	-0.007 	&	\nodata 	&	\nodata 	&	194 	&	15 	&	\nodata 	\\
\enddata
\tablecomments{$^1$ Common duration of LE, ME and HE. \\
$^2$ Burst start time relative to the trigger time. \\
$^3$ Duration of LE, ME and HE, respectively.\\
$^4$ Burst start time of LE, ME and HE relative to the trigger time, respectively,\\
$^5$ Net counts of LE, ME and HE. \\
$^*$ LE data is unavailable due to instrumental effects (e.g., bright Earth, bad events with higher grade). \\
$^+$ LE data and ME data are unavailable due to instrumental effect (e.g., bright Earth, bad events with higher grade). \\
$^{S1}$ The HE data of the burst suffered from data saturation.  \\
$^{S2}$ The LE and HE data of the burst suffered from data saturation.  \\
$^G$ Bursts were also detected by Fermi/GBM. }
\end{deluxetable*}
\end{longrotatetable}

\section*{Acknowledgements}
We thank the anonymous reviewer for very helpful comments and suggestions.
We are grateful to Shri Kulkarni for suggesting the one-month observation after the initial ToO of SGR J1935+2154.
This work is supported by the National Key R\&D Program of China (2021YFA0718500),
the Strategic Priority Research Program on Space Science, the Chinese Academy of Sciences (Grant No. XDA15360300, XDA15052700, XDB23040400),
and the National Natural Science Foundation of China under Grants (No. U1838113, U2038106, U1938201, 12133007, 11961141013).
This work made use of the data from the {\it Insight}-HXMT mission, a project funded by China National Space Administration (CNSA) and the Chinese Academy of Sciences (CAS).


\bibliographystyle{aasjournal}

\end{document}